\documentstyle[11pt,aaspp4,psfig]{article}

\begin{document}
\reversemarginpar

\newcommand{\oii}{{\sc [Oii] }}
\newcommand{\lstar}{L$_*$ }
\newcommand{\sub}[1]{_{\rm #1}}
\newcommand{\cf}[1]{{\cal D}_{\rm #1}}
\newcommand{\nobs}{N^{\rm obs}}
\newcommand{\etal}{{\it et al\/}}

\newcounter{altnum}
\setcounter{altnum}{1}

\renewcommand{\thefootnote}{\fnsymbol{footnote}}

\title{HST IMAGING OF CFRS and LDSS galaxies \\
 --- I: Morphological Properties\footnotemark[\thealtnum]}

\renewcommand{\thefootnote}{\arabic{footnote}}

\author{Jarle Brinchmann\altaffilmark{1}, Roberto
  Abraham\altaffilmark{1}, David Schade\altaffilmark{2}, Laurence
  Tresse\altaffilmark{1}, Richard S.\ Ellis\altaffilmark{1}, Simon
  Lilly\altaffilmark{1}\altaffilmark{,3}, Olivier
  Le F{\`e}vre\altaffilmark{4}\altaffilmark{,6}, Karl
  Glazebrook\altaffilmark{5}, Fran\c{c}ois Hammer\altaffilmark{6}, Matthew
  Colless\altaffilmark{7}, David Crampton\altaffilmark{2}, Tom
  Broadhurst\altaffilmark{8}}

\altaffiltext{1}{Institute of Astronomy,
  Madingley Road, Cambridge CB3 0HA} 
\altaffiltext{2}{Dominion Astrophysical Observatory, Victoria, Canada}
\altaffiltext{3}{Department of Astronomy, University of Toronto,
  Toronto, Canada} 
\altaffiltext{4}{Laboratoire d'Astronomie Spatiale, Traverse du
  Siphon, B.P.8, 13376 Marseille Cedex 12, France}
\altaffiltext{5}{Anglo-Australian Observatory. Siding Spring
  Observatory, Coonabarabran, NSW 2357, Australia}
\altaffiltext{6}{Observatoire de Paris, Section de Meudon, DAEC, 92195
  Meudon Principal Cedex, France}
\altaffiltext{7}{Mt.\ Stromlo and Siding Spring Observatories,
  Australian National University, Weston Creek, ACT 2611, Australia}
\altaffiltext{8}{Astronomy Department, University of California,
  Berkeley, CA 94720, US}
\renewcommand{\thefootnote}{\fnsymbol{footnote}}
\footnotetext[\thealtnum]{Based on observations with the
  NASA/ESA {\it Hubble Space Telescope\/} obtained at the Space
  Telescope Science Institute, which is operated by the Association of
  Universities for Research in Astronomy, Inc., under NASA contract
  NAS 5-26555}
\renewcommand{\thefootnote}{\arabic{footnote}}

\begin{abstract}

  We analyse Hubble Space Telescope images of a complete sample of 341
  galaxies drawn from both the Canada France and Autofib/Low
  Dispersion Survey Spectrograph ground-based redshift surveys. In
  this, the first paper in the series, each galaxy has been
  morphologically classified according to a scheme similar to that
  developed for the Medium Deep Survey. We discuss the reproducibility
  of these classifications and quantify possible biases that may arise
  from various redshift-dependent effects. We then discuss automated
  classifications of the sample and conclude, from several tests, that
  we can expect an apparent migration with redshift to later Hubble
  types that corresponds to a misclassification in our adopted machine
  classification system of $\sim 24\% \pm 11$ of the true ``spirals''
  as ``peculiars'' at a redshift $z\simeq$0.9. After allowing for such
  biases, the redshift distribution for normal spirals, together with
  their luminosity function derived as a function of redshift,
  indicates approximately 1 magnitude of luminosity evolution in
  $B_{AB}$ by $z\simeq 1$. The elliptical sample is too small for
  precise evolutionary constraints.  However, we find a substantial
  increase in the proportion of galaxies with irregular morphology at
  large redshift from $9\%\pm3\%$ for $0.3\le z\le 0.5$ to $32\% \pm
  12\%$ for $0.7 \le z \le 0.9$. These galaxies also appear to be the
  dominant cause of the rapid rise with redshift in the blue
  luminosity density identified in the redshift surveys.  Although
  galaxies with irregular morphology may well comprise a mixture of
  different physical systems and might not correspond to present day
  irregulars, it is clear that the apparently declining abundance and
  luminosities of our distant ``irregulars'' holds an important key to
  understanding recent evolution in the star formation history of
  normal galaxies.

\end{abstract}

\section{Introduction}
\label{sec:intro}
  
The refurbished Hubble Space Telescope (HST) offers the exciting
prospect of addressing the morphological evolution of galaxies
directly from systematic studies of galaxies imaged at various
redshifts. The angular resolution of the Wide Field Planetary Camera 2
(WFPC-2) is 0.1 arcsec which corresponds to a physical scale of less
than $2$ kpc at all redshifts in most popular cosmological
world models (we have assumed $H_0=50$km s$^{-1}$ Mpc$^{-1}$ here and
will adopt this value and $q_0=0.5$ in the rest of the paper).

Considerable progress in understanding the morphological mixture of
the faint galaxy population has already been achieved through the
Medium Deep Survey (\cite{Griffiths-et.al-94};
\cite{Driver-et.al-95}), an extensive imaging programme using
WFPC-2 in parallel mode. Counts of galaxies classed by visual
morphology (\cite{Glazebrook-et.al-95b};
\cite{Driver-et.al-95}) and by other means
(\cite{Abraham-et.al-96}; \cite{Odewahn-Windhorst-Driver-96})
have been compared with model predictions and an apparent excess of
`irregular/peculiar/merger' galaxies is noted when compared to models
based on no evolution. Deeper HST images of smaller areas have been
taken using WFPC-2 in primary mode. Driver et~al. 1995b have
analysed a single deeper pointed exposure of 5.7 hours confirming and
extending the MDS analysis to I=24.5. Abraham et~al (1996b) have
likewise categorised galaxies in the Hubble Deep Field to I=25.  Given
the magnitude limits over which these changes are seen, the above
studies point to fairly recent changes in the morphological
characteristics of the galaxy population.

Similar progress has been made from large systematic ground-based
spectroscopic surveys which serve to delineate the field galaxy
luminosity function (LF) and its evolution out to redshifts
$z\simeq1$.  The $I$-selected Canada France Redshift Survey (CFRS,
\cite{CFRS-I}; \cite{CFRS-II} and references therein) comprises a
complete spectroscopic sample of 591 galaxies in the magnitude range
$17.5\le I_{AB}\le 22.5$ with determined redshifts. The rest-frame
$B_{AB}$-band LF has been determined for various redshifts and
colour-selected components of the population.  Strong evolution with
redshift is found in the luminosity function of the bluer galaxies.
These trends are supported by those from the less deep, but also
extensive, Autofib/Low Dispersion Survey Spectrograph (hereafter LDSS)
redshift survey (\cite{Ellis-et.al-96} and references therein).  This
survey is $b_J$-selected and spans a wide apparent magnitude range
(11.5$<b_J<$24) enabling the {\it shape} of the star-forming component
of the LF to be monitored to z=0.75. Strong evolution is seen in terms
of the space density and luminosity of galaxies with intense star
formation categorised via their \oii emission. Using the same
data Heyl et~al (1997) showed that it is the late {\em spectral\/}
types that dominate the evolution out to $z\simeq 0.5$.  Both surveys
are consistent with a large decline in the luminosity density of star
forming galaxies since a redshift of 1 (\cite{Lilly-et.al-96};
\cite{Ellis-et.al-96}).

The HST imaging and ground-based spectroscopic surveys present
different but complementary views of the evolving galaxy
population since $z\simeq 1$. Indeed, it is tempting to connect the
rapid increase with look-back time in the proportion of galaxies in the
irregular/peculiar/merger category with the strong evolution seen in
the star-forming blue sources in the redshift surveys. However, until
recently, there has been surprisingly little overlap between the wealth
of HST data and the ground-based redshift surveys, largely because of
the mismatch in field size between WFPC-2 and ground-based multi-object
spectrographs.

In this series of papers we plan to remedy this deficiency via a HST
imaging programme of 341 galaxies targetted in either the CFRS or
LDSS surveys. The goals of the study are to employ techniques
developed in the analyses of both the HST and ground-based datasets to
physically understand the origin of the remarkably recent evolutionary
trends identified in the independent datasets.

This paper is concerned with describing the survey parameters and
selection criteria and the techniques used to analyse the HST data.
Further details of the ground-based spectroscopic datasets can be
found in the CFRS and LDSS articles (\cite{CFRS-V};
\cite{Ellis-et.al-96}). By bringing together HST and redshift data for
a large complete sample of distant galaxies, we address in this paper
the question of whether the rapid evolution in the morphologically
peculiar population can be identified with the star-forming blue
sources in the redshift surveys.  As shown below, the correspondence
is convincing, as originally conjectured by the MDS studies.

A plan of the paper follows. In Section~2 we discuss the basic features
of the LDSS and CFRS redshift surveys. As the photometric systems
and the treatment of k-corrections differ between the two surveys,
we compare and align these prior to further analysis. The HST imaging
data is introduced in Section 3 and both visual and automated morphological
classifications are presented. A major question is the extent to which
the apparent morphological type recognised in the HST data is affected
by redshift-dependent biases. Through simulations based on local
multicolour imaging data, we address this point in detail in Section 4
and obtain statistical correction factors which are applied in the subsequent
analyses. Section 5 presents the redshift distributions and luminosity 
functions for three broad morphological classes. We also 
discuss the associated blue luminosity density for each type as a function 
of redshift and interpret this in the context of simple models put 
forward to explain the evolutionary trends found in the redshift survey 
data. Our main conclusions are summarised in Section~6.

Later papers in the series use structural parameters for each HST image
to discuss the physical processes that drive this evolution.  Lilly
et~al. (1997) (Paper II) examines the surface brightness
characteristics of the largest disk galaxies in order to constrain the
extent to which evolution of massive galaxies may be important.
Schade et~al. (1997) (Paper III) addresses the question of evolution
in the number density and photometric properties of the spheroidal
population. Papers II and III both concentrate on quantitative
morphological measures for the galaxies, extending and improving the
analysis in Schade et~al. (1995). Le F\`evre et~al. (1997) (Paper IV)
perform quantitative measures of clustering on small physical scales 
and use this to study the rate of merging as a function of
redshift.

\section{Survey Description}
\label{sec:survey_description}
 
Upon completion of the CFRS and LDSS redshift surveys both CFRS and
LDSS teams independently sought HST WFPC--2 time in Cycle 4. Early
results from these programmes have been described by Schade et~al.
(1995) and Ellis (1995). From 1994, both teams agreed to merge their
efforts and further allocations of HST time to the combined team were
made in Cycles 5 and 6.

In Cycle 4, prior to the merged effort, the strategy adopted by two
teams was somewhat different. Although both teams sought F814W imaging
for morphological classifications, the CFRS team chose to supplement
these images with ones in F450W whereas the LDSS team explored the
visibility of their $b_J$-selected samples in F336W and F218W. Only in
Cycle 5 and 6 did a common strategy emerge based on F814W images
augmented by F450W images for a subset of the targets.

It is important to recognise that the ground-based strategies adopted
by the CFRS and LDSS survey teams differed in several respects.
Foremost the CFRS survey is $I$-selected whereas the LDSS survey is
$b_J$-selected.  The strong differential effects of k-correction with
galaxy type mean that, although the LDSS survey is shallower, it is
more sensitive to the presence of star-forming galaxies that apparently
dominate the evolutionary trends. By contrast, the CFRS survey is less
affected by k-correction effects overall and probes to higher
redshift.  Both teams chose to present their data in terms of
rest-frame blue magnitudes although CFRS presented LFs on a photometric
scale based on B(AB) whereas the LDSS team did so in $b_J$. The
transformation between these two magnitude scales is studied below.
 
Table~1 summarises the HST survey data obtained for this analysis. The
total dataset consists of 341 objects drawn from 25 WFPC--2 HST
frames. 6 fields are drawn from the contiguous `Groth strip'
(\cite{Groth-et.al-94}) imaged by WFPC--2 for which supplementary
spectroscopy of 59 objects was obtained to a magnitude limit identical
to that adopted for the main CFRS survey using LDSS-2 on the 4.2m
William Herschel Telescope.  Galaxies have only been included in the
final catalogue if they have been spectroscopically targeted in either
the CFRS, LDSS or Groth strip surveys.  37 objects were excised as
their HST images are partially or completely obscured by edges of the
WFPC chips or cosmetic defects precluding accurate morphological
analysis. The catalogue contains 7 objects from the Autofib fibre
survey (\cite{Ellis-et.al-96}) as well as 22 LDSS-1 (\cite{LDSS1-III}
and references therein) objects.  The spectroscopic completeness for
these objects is only 62\%, and they will not be discussed further in
the present paper since little is gained in terms of numbers; we will
instead concentrate on galaxies from the CFRS and LDSS-2
(\cite{Glazebrook-et.al-95a}) surveys. The absolute magnitude -- redshift
distributions of the objects from the CFRS and LDSS-2 samples and the
sub-samples for which HST data is available are compared in
Figure~\ref{fig:ldss_vs_cfrs_redshift}. It can be seen that the two
surveys span different regions in this parameter space. The CFRS survey
is particularly useful in probing all classes of luminous galaxies in
the interval 0.5$<z<$1, whereas the LDSS survey is effective in
probing less luminous star-forming galaxies in the interval
0.2$<z<$0.7. The complete catalogue of galaxies comprising the HST
survey is given together with detailed comments in Table~2. 

All observed fields were imaged through the F814W filter which ensures
that a self-consistent and uniform photometric scale can be provided
across both the CFRS and LDSS galaxies. Only the CFRS survey galaxies
currently have reliable ground-based $I_{AB}$ photometry. To achieve a
uniform photometric scale, the raw HST images were processed using the
standard STScI pipeline, and photometry in the F814W system was
performed using the Iraf {\tt apphot} package with 3\arcsec\ diameter
apertures. The zero-point calibration was taken from Holtzman et~al
(1995). The colour transformation between $I_{AB}$ and $I_{F814}$ was
calculated for the CFRS galaxies by interpolating within the observed
$V-I$ colours according to a set of spectral energy distributions
(SEDs); good agreement with the ground based magnitudes was found.

As it was not possible to image {\it all} the ground-based fields in
both surveys within the HST time allocated, the LDSS group initially
selected their fields on the basis of maximising the fraction of
targets for which redshifts had been secured, whereas the CFRS group
originally imaged fields to maximise the fraction of high redshift
objects. After the two groups joined forces, it was agreed that no
particular criterion should be used to select the remaining fields. We
have, retrospectively, verified that the earlier selection criteria
have not unduly weighted the HST survey to an unrepresentative sample
of galaxies.

The completeness in the original ground-based surveys varies from field
to field primarily because of the vagaries of weather at the time of
the original observations. The completeness statistics are given in
Table~3 for the subset of galaxies in the survey that were either in
the CFRS or LDSS-2 redshift surveys. The completeness is only
marginally higher than that appropriate for the parent survey.

Figure~\ref{fig:redshift_contribution} shows the redshift distribution
separately for the CFRS and LDSS-2 objects in the HST survey. Clearly a
greater fraction of the CFRS galaxies have been selected for study with
HST than is the case for the LDSS survey. The median redshift for the
HST-selected CFRS objects, $\langle z\rangle=0.61$, can be compared
with $\langle z\rangle=0.56$ for the entire CFRS ground-based survey.
For the LDSS-2 objects the difference is similarly small --- $\langle
z\rangle=0.46$ in the ground-based survey compared with $\langle
z\rangle=0.43$ in the HST imaged subset. 

Clearly it is important to construct a uniform absolute magnitude scale
across the two surveys. This is an important problem not only because
of observational differences in the photometric selection criteria
used by the two groups but also because of procedural differences used
in estimating the $k$-corrections. To check the photometric
differences, we transformed the LDSS-2 $b_J$ photometry to the
$B_{AB}$ system used by the CFRS.  The required $b_J-B_{AB}$ colour
was found by using the LDSS-2 $b_J-r_F$ to define an SED from which
the colour offset was located by interpolation.  A histogram of the
offsets obtained in this manner is shown in the top panel of
Figure~\ref{fig:magnitude_transforms}, where it is seen that the
offset and scatter are both quite small compared to the bin widths we
will use in discussing evolutionary trends e.g. in the LF.

To investigate systematical differences in the $k$-corrections used
between the two groups, we compared the $k$-corrections derived for a
subset of CFRS galaxies using their set of SEDs located via the $V-I$
colour with those defined similarly based on the LDSS-2 set of SEDs.
The agreement is again surprisingly good indicating a dispersion of
only 0\fm 1 and no systematic offset of significance (see middle panel
Figure~\ref{fig:magnitude_transforms}).

There is one further procedural difference concerning the surveys. In
the case of the CFRS survey, the absolute magnitudes are derived from
isophotal $I\sub{AB}$ magnitudes whereas the LDSS-2 group used
corrected aperture magnitudes. To determine the offsets involved, we
note that the 3\arcsec\ aperture HST magnitudes for the faint CFRS
objects agree well with their ground based isophotal magnitudes. We
therefore measured 3\arcsec\ aperture magnitudes for the LDSS-2
objects on the HST images and calculated absolute magnitudes from
these which were then compared with the published LDSS-2 absolute
magnitudes. This comparison is shown in the bottom panel of
Figure~\ref{fig:magnitude_transforms} and reveals no significant shift
with a dispersion of only $0\fm 2$, enabling us to conclude that the
LDSS-2 corrected aperture magnitudes are closely equivalent to the
CFRS $I\sub{AB}$ isophotal magnitudes. The photometric systems can
thus be aligned by adding the colour term to the LDSS absolute
magnitudes.

In conclusion, within the typical photometric error of $0\fm 2$, there
is no evidence of a serious systematic shift between the two absolute
magnitude scales. For consistency, all absolute magnitudes for
galaxies drawn from {\it both} surveys were calculated with the same
program using $I_{AB}$ isophotal magnitudes for CFRS objects and
3\arcsec\ $I$ aperture magnitudes from HST for the LDSS objects. In
the subsequent analyses, virtually identical results are obtained when
the published absolute magnitudes are used.

\section{Classifications}
\label{sec:classifications}

To physically interpret the evolution of the galaxy LF as delineated
by the original ground-based redshift surveys, both the CFRS and LDSS
analyses sub-divided the samples on the basis of spectroscopic and
photometric classes. CFRS analyses of the LF based on rest-frame color
found luminosity evolution to be stronger for galaxies bluer than a
Sbc. LDSS samples selected according to the rest-frame equivalent
width of the [O II] 3727 line found that the evolutionary
trends arose almost exclusively from galaxies with strong
emission-lines.

The availability of the HST data for 341 galaxies allows us to
investigate these changes in more detail. Ideally galaxies should be
classified according to a label which is not modified by any of the
physical processes responsible for the evolution. Part of the
difficulty with colour and emission line strength is that populations
defined according to these criteria may well be transient, and thus
detailed comparisons of luminosities and volume densities at various
redshifts will be confused.

Although the same criticisms can no doubt be applied to galaxy
morphology (\cite{White-96}), the morphologically-dependent number
magnitude counts derived from the Medium Deep Survey and Hubble Deep
Field (\cite{Glazebrook-et.al-95b}; \cite{Driver-et.al-95};
\cite{Abraham-et.al-96}) raise important questions concerning the
apparent rapid evolution of the irregular/peculiar/merger galaxies in
comparison with the slower trends noted for the spheroidal and regular
spiral classes. How do these classifications map onto the redshift
survey data plane? In this section we discuss the various ways in
which we have classified the galaxy morphologies taking care to note
these uncertainties and systematic changes that may occur because our
survey samples galaxies over a large range of redshifts.

\subsection{Visual Classifiations}
\label{sec:visual_class}

The first technique we used to investigate the morphological
characteristics of our sample follows the visual approach adopted by
the Medium Deep Survey team.  Following the precepts discussed
by Glazebrook et~al. (1995b), three of us (RSE, SJL, OLF) have
classified all the galaxies by eye according to a scheme illustrated
in Figure~\ref{fig:examples_of_morphology}. The scheme we have adopted
here differs slightly from that utilised by the MDS team in that we
decided to separate compact objects with faint extensions (so-called
``tadpoles'') from the irregular/merger/peculiar and compact objects.
This was done for objects for which classifications were either
compact or peculiar and where no consensus could be reached.
 
An intercomparison of the eyeball classifications between the 3
observers is shown in Figure~\ref{fig:correlations} and indicates a
scatter of around 1.2 classes. This is similar to the scatter found by
the MDS team (\cite{Abraham-et.al-96b}) at the same magnitude limit.
The individual classifications were then merged by taking  the
median value of the different classifiers. The final number of objects
in various classes is also indicated in
Figure~\ref{fig:examples_of_morphology}.  The resulting $M_B-z$
diagram is shown in Figure~\ref{fig:mbz_figure} with
the fractional redshift distributions of the various morphological
classes inlaid.
 
In order to determine whether the morphological mixture is robustly
estimated, we can compare the morphologically segregated $N(m)$ for
our sample with that for the MDS survey (\cite{Abraham-et.al-96};
\cite{Glazebrook-et.al-95b}). Since the MDS survey is $I$-selected, we
can only compare it with that subset of our galaxies drawn from the
CFRS survey. As the CFRS survey did not target all objects lying
between the photometric limits, we must correct for those objects that
are in the HST field within the magnitude limits which were not
targetted spectroscopically. We have done this by multiplying the
counts in each field with the ratio of photometrically to
spectroscopically observed objects for each HST frame, taken from the
CFRS survey. The resulting counts are shown in
Figure~\ref{fig:nm_variation}. It can be seen that there is an
apparent lack of bright irregular galaxies. This is in part because
the bright irregulars by coincidence happen to be in fields with high
completeness.  By assigning the bright irregulars to random fields we
find the difference between our irregular counts and the MDS counts is
generally less than 2$\sigma$ and we do not consider this as a
potential problem, as we will see the abundance of low redshift
irregulars is almost exactly as expected from the local luminosity
function (see Section~\ref{sec:redshift_distribution}). We find an
integrated count from $I_{AB}=17.5$ to $I_{AB}=22.5$ from our survey
of 394 whereas that expected from the MDS survey is 409, i.e.  in
close agreement.  The field to field variation of the morphological
composition is also satisfactorily constant to within the
uncertainties.
 
\subsection{Automated Classifications}
\label{sec:automated_class}

A major difficulty with visual classifiers is their subjective nature
(\cite{Naim-et.al-95}). Accordingly, as an objective route forward, we
have also performed machine-based classifications using the procedures
adopted by Abraham et al.\ (1994,1996b). The technique is based on
measurements of a central concentration index, $C$, and a rotational
asymmetry factor, $A$. The first of these parameters tracks the
bulge-to-disk ratio, while the second traces the degree of
irregularity. At low redshift, the positions of galaxies on the
log($A$) vs log($C$) plane can be used to distinguish between
early-type systems, spirals earlier than type Sd, late-type
spirals and irregulars/peculiars/mergers. As will be shown below, at
higher redshifts these classes can also be distinguished in principle
using a $A-C$ diagram. However, uncertainties arise because galaxies
of a given type may move into a  region of the $A-C$ plane
occupied by a different class because of redshift-dependent effects.
These biases must be carefully studied before quantitative comparisons
can be made over a range in redshift. We defer a discussion of these
studies until the next section.

The starting point for measuring the two parameters is a ``segmented''
galaxy image, constructed by isolating those pixels that lie above a
surface brightness threshold $N\sigma$ above the sky brightness, where
$\sigma$ is the sky variance and $N$ is a constant, typically 1.5.
For the present work (see discussion below and in Appendix A), we have
chosen $\sigma$ so the measurements go to a uniform limiting surface
brightness.  The $C$ parameter represents the ratio of light within an
inner and outer elliptical aperture determined from the
sky-subtracted, intensity-weighted, second order moment of the
resulting image. The major and minor axes of the outer aperture are
normalized so that the total area within the ellipse is the isophotal
area of the galaxy. The inner aperture is defined by scaling these
axes down by a linear factor of three.

Whereas this definition of central concentration is adequate for local
galaxies, the value determined unfortunately depends on redshift since
the threshold is defined relative to the sky. Thus less of the galaxy
is sampled at high redshift because of cosmological dimming. Measuring
$C$ to a fixed rest-frame surface brightness isophote is not really
practical so it is necessary to consider how to correct $C$ for this
effect. A procedure to do this is discussed in
Appendix~\ref{sec:c_correction}, and in the rest of the paper all $C$
values have been corrected using the minimal correction defined in
equation~(\ref{eqn:c_correction}).

The rotational asymmetry parameter $A$ is defined via:
\begin{equation}
\label{eqn:A_calc}
A = {\sum_{ij}\left| {{I_{ij} - I^R_{ij}}}\right | \over \sum_{ij}
  I_{ij}}   - k_{A}.
\end{equation}

\noindent where $I_{ij}$ is the intensity in pixel $(i,j)$, and
$I^R_{ij}$ is the corresponding intensity after image rotation by
180$^\circ$ about the centroid of the segmented galaxy image. The
$k_A$ term in equation~(\ref{eqn:A_calc}) is a small correction
accounting for signal introduced into $A$ by noise in the sky
background; $k_A$ is determined by measuring the asymmetry within a
rotated and self-subtracted region of sky equal in area to that of the
galaxy being analysed.

The measurement of these parameters is only practical for that subset
of galaxy images with more than 64 contiguous WFPC--2 pixels above the
surface brightness threshold (see also the discussion
in~\cite{Abraham-et.al-94}). 25 galaxies are too compact for reliable
classification via this approach.  For these objects, only the visual
estimates are available. The upper panel of Figure~\ref{fig:ac_plot}
shows the corrected $A$ and $C$ distributions for our sample. Dashed
lines define three morphological bins (ellipticals/ spirals/peculiars)
drawn on the basis of similar measurements made on a local sample of
galaxies of known morphology (\cite{Frei-et.al-96}).  Detailed
consideration of this local dataset is deferred to the next section.
To avoid confusion with the visual classifications, we will refer to
the automated classes as $AC$-ellipticals ($AC-E$), $AC$-spirals
($AC-S$) and $AC$-peculiars ($AC-P$). The angle from the intersection
point of the dashed line in the upper panel of
Figure~\ref{fig:ac_plot} to each point in the AC plane, defines a
continuous classification angle, $\Theta$. The correlation between
this angle and the eyeball classifications is shown in the lower panel
of the figure. Overall the agreement is satisfactory but a sizeable
scatter is apparent. Note in particular that the angle does not
distinguish well between early-type spirals and E/S0 galaxies.

\section{Redshift-dependent Biases}
\label{sec:bandshifting}

High $z$ galaxies imaged by the HST differ in appearance from their
local counterparts because of their reduced apparent size and sampling
characteristics, a lower signal-to-noise and reduced surface
brightness with respect to the sky background, and a shift in the
rest-wavelength of the observations.  The latter term we will refer to
as `bandpass shifting'.  These effects will combine to give some
uncertainty in the morphological classification of galaxies, generally
in the sense of shifting objects to apparently later Hubble types.

Glazebrook et~al (1995b) attempted to address this in the context of
their visual classification procedure via a blind classification of
local galaxies whose appearance was carefully simulated as viewed at a
redshift $z=0.7$. For machine classifications this has been addressed
to some extent by Abraham et~al. (1996b) using the parameters $A$ and
$C$. A benefit of working in the framework of $A$ and $C$
classifications is that it enables more quantitative statements about
these biases.

In this paper we will extend the discussion begun by Abraham et al
utilising the known redshifts of our sample to correct for these
redshift-dependent biases. To do this we have relied on extensive
simulations based on a set of multi-colour CCD images of local galaxies
from Frei et~al (1996). Our approach will be to assess (in various
ways) the extent to which the A\&C parameters for local galaxies of
known type are likely to shift when they are placed at larger
redshift.  Statistically, over a number of galaxies, we then compare
the distribution of classes, as viewed at a given redshift, with the
intrinsic value at $z\approx$0 to determine a `misclassification
fraction' for each type which is a function of redshift. The method
assumes that the Frei et~al galaxies of a given type are
representative. Note, in particular, that we do not require an accurate
sampling of the local morphological mix.  Indeed, in principle, only a
few representative galaxies of each type are required. Our goal and
procedure differs from the synthetic creation of faint no-evolution
samples (see eg.\ Bouwens, Broadhurst \& Silk 1997) for which the Frei
et~al sample is not well suited.

\subsection{The Frei  et al Calibration Sample}

The Frei et~al sample consists of 82 galaxies with $B_J$ and $R$
CCD images, uniform in quality with foreground stars removed. It is 
important to note, however, that the Frei et al galaxies were not 
chosen with the intention of sampling the luminosity function of local 
systems uniformly. Indeed, Frei  et al chose galaxies that are 
(a) bright, (b) have well-resolved morphological structures, and 
(c) span a wide range of Hubble system classification classes. It 
is therefore important to assess whether this sample is appropriate
for calibrating redshift-dependent morphological trends.

The {\em absolute magnitude distribution} of the Frei  et al
sample is shown in Figure~\ref{fig:bobs_figure}. The plot is based on
$M_B$ data published in the Revised Shapely Ames Catalog (RSA). We
have applied an additional $0\fm $5 shift to take account of
corrections applied for internal extinction in the RSA which we ignore
in the CFRS+LDSS samples. We also indicate characteristic $M^\star$
values for local elliptical and Scd galaxies (\cite{Marzke-et.al-94}).
The Frei  et al data peaks near $M^\star$ and, as expected, is
deficient in systems $\sim 2$ mag or more fainter than $M^\star$. This
could be a drawback for our purposes, depending on how the
morphological biases are corrected. If, for example, we correct
morphologies back to those appropriate for the rest-frame $I$-band, a
proper match between the CFRS/LDSS and Frei et~al luminosity
distributions is more important at high redshift. In this case, the
majority of our survey galaxies beyond $z\simeq 0.3$ are drawn from
within two magnitudes of $M^\star$. In Figure~\ref{fig:bobs_figure} we
show lines corresponding to the apparent magnitude limits for the CFRS
survey based on $k$-corrections for early and late-type systems.
Clearly the underluminous galaxies which are deficient in the Frei
sample are undetectable. For our approach this leads to a low number
of objects, and hence a large statistical uncertainty. But since we
cover the whole range of morphological types, we do not introduce any
systematic biases.

One might worry that selecting the galaxies to be nice and regular
looking might lead us to infer less bandshifting than is
observed. However, we do not think this is a major problem for our
approach, since the comparison with the low-redshift data effectively
takes out the actual values of $A$ and $C$. Thus, although the Frei
{\it et al} sample is useful for estimating the apparent shift in
morphology with redshift for regular objects of suitable luminosity,
it could be improved. Future imaging surveys of local systems should
aim to sample fairly the {\em distribution} of morphological types
within both the luminosity selection criteria of the survey and the
regions on the $A$ vs. $C$ diagram. Such datasets would also allow
compilation of no- and mild-evolution simulated datasets for
comparison with high redshift data (see also Bouwens et al.\ 1997).

\subsection{Wavelength-dependent Trends}
\label{sec:redshift.biases}

The simplest test we can perform is to measure the $A$ and $C$ parameters
defined earlier in both the $R$ and $B_J$-band images of the Frei 
{\it et al\/} galaxies. Since our faint HST images are taken with the 
F814W filter, we effectively see the $R$ band at a redshift of 0.2 and 
the $B_J$ band at a redshift of 0.87. The shift in $A$ and $C$ across
the Frei {\it et al} sample is thus a crude but simple measure of the
shift arising from the bandpass effect over much of the redshift range
sampled.

In the top panel of Figure~\ref{fig:deltaa_deltac} we plot the change 
in asymmetry $A$ observed using the $R$ and $B_J$ images of the same 
local galaxy. Late-type systems are denoted by filled circles and, as 
expected, there is a clear trend for such systems to have larger asymmetry 
at shorter wavelengths where the star-formation signatures are more
irregular. However, in quantitative detail, the size of the effect is 
quite small. The corresponding shift in concentration $C$,  shown in 
the bottom panel, is somewhat larger. The change in $C$ required
for a galaxy to cross the $AC$-peculiar and $AC$-elliptical boundary is
also indicated on the figure; for the $AC$-peculiars this is strongly
dependent on asymmetry.  

This simple comparison indicates that only a small fraction of the
Frei {\it et al} galaxies would cross into the areas defining
different morphological types. However, the comparison is crude and
takes no account of more complex sampling and noise effects.

\subsection{Results from Detailed Simulations}
\label{sec:redshifting_results}

To quantify the redshift biases more precisely, we decided to incorporate
each of the effects that combine to give the final HST appearance.  
The method adopted is based on that described in more detail by 
Abraham et~al. (1997) and can be summarised as follows:
\begin{itemize}
\item For each pixel we calculated the $B_J-R$ colours; to avoid edge
  effects we smoothed the images slightly before calculating the
  colours.
\item The pixel $B_J-R$ colours were then used to select an SED to each
  pixel (as for the integrated colours discussed earlier). This SED 
  was then used to determine the k-correction applicable to each pixel.
\item The final step is to rebin the image, using the known and wanted
  redshift, applying $(1+z)^4$ surface brightness dimming and adding noise
  corresponding to the characteristics of WFPC--2 for our mean exposure
  time.
\end{itemize}

For each redshift we only analysed those redshifted galaxies that
would have been selected into either the CFRS or LDSS--2 redshift
surveys.  Beyond $z\simeq0.9$, however, only a few of the Frei et~al
galaxies would be seen; this is because the bright galaxies in the
Frei sample tend to be Sab galaxies whose k-corrections are
substantial when the 4000\AA\ break enters the I-filter (this is also
indicated in Figure~9). This leads to a small number of galaxies, and
hence an increase in the statistical uncertainties. To rectify this
problem, we adopted an alternative approach.  We selected the Frei
{\it et al\/} galaxies that would have had $I_{AB} <23.50$, ie.\ one
magnitude too faint, and brightened these galaxies so that they fell
within our selection criteria. This led to an average brightening of
$\langle M_B\rangle \approx 1.5$ magnitudes, and kept $M_B\ge -22.5$.
This enabled us to get a satisfactory number of objects without
introducing any over-luminous objects. Provided the morphological
characteristics do not vary significantly over this small luminosity
range, this should not affect our conclusions.

The aim of the detailed simulations is to determine that fraction
of a given type of Frei {\it et al} galaxy which {\it appears} to be 
of a different morphological type (as measured by $A$ and $C$) at a
chosen redshift $z$. For convenience we will denote the number of objects 
in a given category by $N$ with superscript 'obs' for the observed number 
and no superscript for the true number in that class. We connect these
two numbers via a drift coefficient, $\cf{XY}$ which characterises the
drift from category X to category Y, defined as
\begin{equation}
  \label{eq:definition_of_cf}
  \cf{XY}=\frac{N_{X\to Y}}{N_X},
\end{equation}
where $N_{X\to Y}$ is the number of objects of class $X$ that are
classed as $Y$ at a higher redshift. From the simulations we can
estimate $\cf{XY}$ for various redshifts.

This methodology is similar to that of the k-correction applied to
convert an observed magnitude into a rest-frame value. The ultimate
aim here is to recover the {\it rest-frame morphology}. Although the
analogy fails in detail because of the added complications of
distance-dependent resampling and surface brightness dimming, our
simulations indicate that these are second order effects. If the
central concentration is corrected for surface brightness effects as
described in section~\ref{sec:automated_class}, the dominant cause of
a migration to a different morphological type is the pixel-by-pixel
k-correction.

The analogy with the k-correction suggests two approaches for analysing
our data. Given we have $R$-band images of the Frei galaxies and the
local morphological studies are done in $B$, it seems natural to
correct our HST morphologies to those appropriate for the rest-frame
$B_J$-band. As we observe this rest wavelength directly at $z\approx
0.9$, the corrections would be small at high redshift but larger at low
redshift. A problem with this approach is that our benchmark sample is
not ideal for studying the morphological shifts at the low luminosities
appropriate for the nearby objects, due to the low number of faint
galaxies. 

The alternative approach is to correct our HST morphologies to rest
frame $R$ assuming, as seems reasonable, that the morphology changes
little between rest-frame $I$ and $R$. In this case, nearby,
intrinsically faint galaxies, need no correction and although a larger
correction is needed at high redshifts, in this case the Frei {\it et
  al\/} sample is well matched in luminosity. Neither approach is
perfect, but we consider the latter to be more reliable until larger,
multiband local samples are available. We will therefore concentrate
on applying corrections to rest-frame $R$ morphologies (although, for
completeness, we tabulate drift coefficients appropriate for
rest-frame $B_J$).

The results of this exercise are summarised in Table~4. The numbers
here are consistent with the earlier discussion of the change in $C$
from the $R$ to $B_J$-band images but we consider the results to be
more reliable given the more detailed treatment of the effects of
sampling and background noise. In particular, the drift coefficients
calculated from $R$ to $B_J$ band images agree excellently with the
ones found here. The drift coefficients not listed were all found to
be zero in the simulations.

We can relate the observed number of objects in class $X$ to the
true number through
\begin{equation}
  \label{eqn:drift_array}
  N^{obs}_X=N_X+\sum_{Y\ne X}^{\mbox{Classes}} N_Y \cf{YX}-
   N_X \sum_{Y\ne X}^{\mbox{Classes}} \cf{XY}.
\end{equation}
Equations~(\ref{eqn:drift_array}) can be readily solved given the
observed number of galaxies. 

Broadly speaking, there are two dominant effects. First there is an
apparent migration from $AC$-spirals to $AC$-peculiars if we classify
galaxies from observed $I$-band images. This bias is expected to occur
also for eyeball classification. At $z=0.7$ we can expect only 13\% of
the $AC$-spirals to be misclassified as $AC$-peculiars, whereas by
$z=0.9$ the misclassed fraction grows to 24\%. Presumably this trend
continues at higher redshift. Due to the restrictions in the local
dataset discussed above, we have not extended the simulations beyond
$z=0.9$.  U-band imaging would be highly desirable to quantify
bandshifting beyond $z=1$ (see also~\cite{Hibbard-Vacca-97})

In addition there is a strong trend for {\it AC\/}-ellipticals to be
migrate into the {\it AC\/}-spiral category. The net result is
somewhat less clear as there is also a drift in the opposite sense
(from {\it AC-S\/} into {\it AC-E\/}) due to random measurement errors
on $C$ where the boundary is nearly vertical.

From the errors in Table~4 it is evident that the drift coefficients
are uncertain. A larger sample of calibrating galaxies is clearly
required to make more precise statements about the drift coefficients.
Nevertheless, the principal conclusion of the simulations is a
reasonably precise measure of the proportion of high redshift
$AC$-peculiars which are likely to be genuine spiral galaxies. The
mixture of regular spirals and spheroidal galaxies should be
faithfully observed with HST out to redshifts $z\simeq$1, however with
a likely loss of {\it AC-E\/} at high redshift which might be able to
make up for the loss of $AC$-spirals to the $AC-P$ category.  A
montage of the $AC$-peculiars sorted by redshift and rest-frame [O II]
equivalent width is shown in Figure~\ref{fig:ac_irregulars}.

\section{Analysis}
\label{sec:analysis}

The first question we address relates to luminosity evolution as a
function of galaxy class inferred independently from the MDS and
ground-based surveys. We wish to understand these results in the
context of the morphologically-segregated redshift distributions and
luminosity functions (LFs) now available to us. Such results are of
considerable interest, as recent semi-analytical models
(\cite{Baugh-et.al-96a}; \cite{Shimasaku-Fukugita-97}) already claim
to reproduce observed trends in the global star formation history
based on ground-based redshift surveys and Lyman limit selected
samples in the Hubble Deep Field (\cite{Madau-97}). However, a physical
understanding of these trends in the context of these models demands a
more detailed comparison such as is now possible for each of the
various morphological types. Although hierarchical assembly may
transform late-type systems into more regular spheroidal and disk
galaxies, in principle these effects can be incorporated in such
models.

In discussing the observational results, as mentioned in
Section~\ref{sec:survey_description}, we will concentrate only on that
subset of the survey containing objects from the LDSS-2 and CFRS surveys.
This provides a total of 249 galaxies with secure redshifts $z<1.2$. In
order to implement the quantitative results on redshift-dependent biases
from the previous section, we will restrict discussion to types based on
the $AC$-classifications. For 12 objects whose isophotal area is less
than 64 pixels, or whose images lie within the planetary camera, the
$A$ and $C$ measurements were replaced with eyeball classifications.
One galaxy among the 249 was classed as compact. Given the
uncertainty associated with dealing with such an object, we increased
the error bar in the relevant redshift range accordingly. In total,
there are 24 spectroscopically-confirmed stars, four QSOs with $z>1.2$
and 35 objects with uncertain redshift (note=1) or no redshift estimate
at all. The failures have been ignored in the analysis (and would
not change the main conclusions below if included).

\subsection{Redshift distributions}
\label{sec:redshift_distribution}

It will be helpful in discussing the observed type-dependent redshift
distributions, $N(z)$, to have no-evolution predictions based on local
LFs. To make these predictions we adopted type-dependent Schechter
LFs listed in Table~5. For the spirals and ellipticals, these are
updated versions of those used by Glazebrook et~al. (1995b) with
$\phi^\star$ adjusted to give the observed fractions given by Shanks
et~al. (1984) in their $b_J<16.7$ sample. For the irregular/peculiar
galaxies we have adopted the late-type/irregular LF given by Marzke
et~al. (1994). Using these local LFs and the $k$-corrections discussed
earlier, we calculate $N(z)$ for the LDSS2 and CFRS galaxies using the
selection criteria and areas listed in Table~3.  For the spiral and
spheroidal galaxies, we are primarily interested in the luminosity
scales at the bright end and so even though there are considerable
uncertainties in the local LFs, the predictions based upon them are
nonetheless a useful guide.

As well as indicating the expected $N(z)$ distributions for the
combined CFRS and LDSS-2 magnitude limits, appropriately weighted for
the sample sizes involved, we also calculate the effect on the
distribution of a luminosity evolution equivalent to a linear shift in
$M^\star$ with redshift that amounts to one magnitude at $z=1.0$. We
will refer to this prediction as `mild evolution'.  We then determined
the observed number of objects as a function of $AC$-class in each
redshift bin fully incorporating the effects of morphological bias as
determined in the previous section using
equation~(\ref{eqn:drift_array}). Since the local morphological mix for
the luminosity functions is based on $B$-band morphologies, we use the
$\cf{XY}$ values for correction to $B_J$-band morphologies, taken from
Table~4. The $N(z)$ distributions for the various types of $AC$ class
are compared with the no-evolution and mild evolution predictions in
Figure~\ref{fig:nz_distributions}.

Figure~\ref{fig:nz_distributions} shows that, taking into account the
biases discussed above, the number of high $z$ ellipticals and spirals
is broadly consistent with the expectations based on $N(m)$ counts.
Given the limited size of our survey, the possibility of
incompleteness and, especially, the reliance that has to be made on
the local LF normalisation in such comparisons, it is difficult to
make a precise statement on the extent of any luminosity evolution. A
K-S test applied to the elliptical $N(z)$ is unable to distinguish
between the no evolution and mild evolutionary predictions. The size
of the spiral sample is larger but precise conclusions are difficult
to obtain because of the number of spirals without redshifts.  Even
so, the mild evolutionary case is preferred: the no evolution $N(z)$
is rejected at the 97\% level and this confidence level would be
stronger if the failures are at high redshift as is most likely the
case. The drop in numbers past $z=1$ is most likely due to
incompleteness in the redshift determinations (see
also~\cite{Cowie-et.al-96}).

The most convincing result apparent from
Figure~\ref{fig:nz_distributions} is the very considerable excess
population of $AC$-peculiars beyond $z\simeq$0.4 which cannot be
explained through residual uncertainties in the misclassification
fraction as applied to the regular spirals.  The figure shows quite
clearly how the large excess population recognised from early
ground-based studies arises primarily from these sources. Note also
that the {\it form} of the redshift distribution is skewed to high
redshift and thus a mistaken normalisation in the local LF would not
be helpful in reconciling the data with the simple model predictions.
The result is highly suggestive of an evolutionary effect.  Although
it is important to remember that the $AC$-peculiar category may
include a variety of physical types (see in particular the discussion
of $I-K$ colours of these objects in~\cite{Glazebrook-et.al-97}) and
that, in principle, morphology itself may be a transient phenomenon,
understanding the strong redshift dependence in the abundance of the
$AC$-peculiars is clearly a crucial goal for making progress.

\subsection{Luminosity functions}
\label{sec:lfs}

The previous section exploits one aspect of the $M_B-z$ plane to
discuss the properties of the survey galaxies, but a disadvantage in
the interpretation of Figure~\ref{fig:nz_distributions} is the need to
assume a local LF and particularly its normalisation.  In order to
understand {\it how} the population of $AC$-peculiars evolves so
dramatically to provide the excess population, and to complement the
study of {\it AC-S \/} and {\it AC-E\/}, it is therefore valuable to
consider the form of the luminosity function at various redshifts.
This can be calculated for the HST survey galaxies using a
$V\sub{max}$ formalism.

As the CFRS and LDSS surveys have quite different selection criteria,
we utilise an approach similar to that adopted
by Ellis et~al. (1996).  For each galaxy in the survey,
$V\sub{max}$ was calculated using
\begin{displaymath}
  V\sub{max,{\it i}}=\sum_j^{\rm Surveys} V_{ij},
\end{displaymath}
where $V_{ij}$ is the normal accessible volume of galaxy $i$ in survey
$j$. To cope with the different selection criteria in the two surveys,
estimates of the $b_J$ magnitudes of the CFRS galaxies and $I\sub{AB}$
magnitudes of the LDSS-2 galaxies are required. For the
LDSS-2 objects we used the HST 3\arcsec\ magnitudes and converted
these to $I\sub{AB}$ following methods discussed above. For the CFRS
objects we calculated $b_J$ magnitude synthetically from $V$
photometry similarly. The method was then checked for a subset for
which $B_{AB}$ photometry was available and a good agreement was found. For
consistency, colour terms were calculated using the same interpolated
SEDs that were used for the calculation of the absolute magnitudes.
The effective areas were taken from Table~3.
The LFs estimated in this way are plotted for three redshift bins in
Figure~\ref{fig:luminosity_functions}. Error bars were determined using
bootstrap resampling techniques.

The redshift biases determined in Section~\ref{sec:redshifting_results}
were incorporated in a simple fashion. We only took account of the
shift of $AC$-spirals into the $AC$-peculiar category in the highest
redshift bin. We then assigned 24\% of the $AC$-peculiar to the $AC$-spiral
class, for each bootstrap repetition. The effects are not significant 
in the following discussion.

A major disadvantage of multi-object redshift surveys with narrow
ranges in apparent magnitude is that they provide very little overlap
in LF across the different redshift bins. Nonetheless, the figure
shows that the LFs for $AC$-ellipticals do not exhibit {\em strong\/}
evolution. The $AC$-spiral LFs show a shift of $\approx 1$ magnitude
to $z=1$ but the greater component occurs between the two lower
redshift intervals. The LFs for the $AC$-peculiars clearly does show
substantial evolution from $z=0$ to $z=1$, in the sense of either a
dramatic brightening with redshift or a substantial increase in the
volume density of luminous examples. It is difficult to quantify this
evolution precisely because of the lack of overlap between the various
redshift bins. In particular we do not observe {\em any\/}
luminous {\it AC\/}-peculiar galaxies in the lower redshift bin, even
though they should be detected given our magnitude limits.  However,
as with the redshift distributions, it seems that the apparently
declining population of luminous $AC$-peculiars is the dominant effect
(as originally proposed by the MDS team), but with a significant
contribution also from the {\it AC\/}-spirals.

\subsection{Luminosity densities}
\label{sec:lumn_density}

We now attempt to quantify the extent to which the evolution we have
found for the $AC$-peculiars contributes to that observed for the
overall population. Numerous authors (\cite{Lilly-et.al-96};
\cite{Madau-et.al-96}; \cite{Madau-97}) have expressed the overall
evolution of the galaxy population in terms of a rest-frame luminosity
density or a volume-averaged star formation rate as a function of
redshift. Over the redshift range 0$<z<$1, these articles interpret
the observations in terms of a substantial decline in star-formation
activity at recent times. A crucial question is whether virtually all
of this decline arises from the rapidly-evolving irregular component
discussed above.

We can address this by examining the rest-frame $B(AB)$ luminosity
density of the galaxies in our HST survey as a function of
morphological type. The results of this exercise are shown in
Figure~\ref{fig:lum_density} where the errors bars are obtained
through bootstrap resampling as before. To correct for the fact that
we observe only a limited magnitude range, we fit Schechter functions
to the luminosity functions in Figure~\ref{fig:luminosity_functions},
with the faint end slope fixed to $\alpha=-0.5$ for the AC-E and
$\alpha=-1$ for AC-S and AC-P. It is clear from that figure that this
leads to considerable uncertainties. In particular the fit for AC-P in
the low redshift bin is too uncertain to be useful. The corrected
values have been plotted as open symbols in
Figure~\ref{fig:lum_density}.

Since the correction for $AC$-peculiars at low redshift is unknown,
one could argue that their rapid rise in the blue light density is
just an artifact of the limited magnitude range sampled at low
redshift. However, it is interesting to note from
Figure~\ref{fig:luminosity_functions} that we do not see any bright
$AC$-peculiars, {\em even though they would lie within the selection
criteria}. This constrains the luminosity function for this region,
and a conservative upper limit to the correction factor is 1.5. Thus
we would argue that the rapidly-increasing contribution to the blue
light by the $AC$-peculiars is a robust result. Over $z\simeq$0.3-0.9
this class provides an order of magnitude increase in the detected
luminosity density in a given magnitude limited sample, consistent
with their dominant effect of the HST galaxy counts over the range
detected by the surveys. The {\it AC\/}-spirals contribute smaller
amount to the overall evolution. Their contribution is not so
evident in the HST galaxy counts, and can be attributed to our being
able to correct for redshift-dependent effects in the classifications.
At high redshift, galaxies would be assigned too low $C$ values, due
to surface brightness dimming. The correction in Appendix~A rectifies
for this and hence demonstrates the importance of allowing for redshift
dependent effects.

\subsection{The physical nature of the star-forming galaxies} 
\label{sec:star_formation}

Given that galaxies of irregular morphology are rapidly evolving with 
redshift in their abundance and/or mean luminosity (and hence detected 
luminosity density), and that this trend appears to dominate the evolution 
we see, at least in the context of galaxy counts selected according to 
apparent magnitude, these remarkable systems clearly hold the key 
to understanding the rapid demise in star formation activity since $z\simeq$1. 

A basic question we must address is whether this category represents a
single type of object evolving in isolation, perhaps fading in
luminosity after an energetic burst of activity, or are we witnessing
the gradual transformation of galaxies, rendered irregular in form
simply by virtue of their enhanced star formation, into more regular
systems? 
Progress might be made on this question if it could robustly be
demonstrated that the rapid evolution seen in the $AC$-peculiars does
or does not occur at the expense of accompanying changes in the regular
spirals. That the high redshift volume density of spirals is
consistent with local estimates does imply, at first sight, that the
$AC$-peculiars might evolve independently of the more modest changes
that affect the spirals. Of course, a decline in star formation
activity in a long-lived class e.g. the spirals, will most likely be
accompanied by a drop in luminosity and such transformations may not
always occur above the detection limits of the survey so it is
difficult to use this argument with confidence. Also, conclusions
derived on the basis of number conservation are uncertain in our case
given our small sample size and the remaining effects of redshift
incompleteness.  Greater progress may be possible in constraining the
growth of the spiral population by attempting to examine a
well-defined subset taking due care to allow for the effects of size
and surface brightness. This is the approach adopted in Paper
II.

Finally, as we have emphasised, in morphological terms, the
$AC$-peculiars represent a mixture of very different objects
(Figure~\ref{fig:ac_irregulars}). The category includes some very
late-type spirals whose asymmetries and central concentrations place
them in the $AC$ plane normally occupied by low $z$ irregulars, double
systems most likely in the act of merging and other peculiar systems
which defy classifications in the normal Hubble sequence.  Paper IV in
this series examines the structure properties and merger statistics in
an attempt to quantify the dominant sub-processes that may drive the
evolution in this class. 

Examination of Figure~11 suggests that a significant fraction of the 
$AC$-peculiars with high \oii\ equivalent widths appear to be rather 
compact. These could be the more extreme examples of the star-forming 
population which dominates the evolutionary trends discussed above.
In this context, it is interesting to consider Guzman et~al's (1997)
claim that a considerable fraction of the star formation activity seen 
at high redshift occurs in ``compact'' galaxies. The compact galaxies 
in their study were selected within the Hubble Deep Field flanking fields 
on the basis of apparent magnitude, angular size and average surface 
brightness. Although their sample is somewhat smaller than that analysed
here, both their magnitude and surface brightness limits are generally 
fainter. However, our HST exposure times are generally longer.

A key point in understanding Guzman et al's result in the context of
this paper where the bulk of the evolution occurs in galaxies of
irregular morphology is the precise definition of a ``compact''
galaxy.  Guzman et al adopted a half light radius $r_{1/2}$ of less
than or equal to 0.5 arcsec as well as a surface brightness selection
criterion  (\cite{Phillips-et.al-97}). We can ask how many of our
regular and irregular galaxies would fulfill this compactness
criterion at various redshifts. From our data we find that as many as
37\% of the $AC$-spirals and 26\% of $AC$-peculiars beyond z=0.5 have
$r_{1/2}<0.5\arcsec$ and $\mu_{F814} > 22.24$. 

The comparison can continue by examining the volume-averaged star formation 
rate (SFR) for the galaxies detected in our survey based, as in Guzman
et al, on the measurements of the equivalent width (EW) of \oii\ and the
relation between EW\oii and SFR from Guzman et~al. (1997),
\begin{equation}
  \label{eqn:sfr_guzman}
  {\rm SFR} (M_{\odot} \mbox{yr}^{-1}) \approx 2.5\times 10^{-12} \times
  10^{-0.4 ( M_B - M_{B_\odot})}  \mbox{EW}_{\mbox{\oii}}.
\end{equation}
This corresponds to $(U-V)_{AB}=0.7$ in the simple model of~Hammer
et~al. (1997) --- a colour which is representative for the galaxies in
the higher redshift interval.

Adopting the same redshift intervals as for the luminosity density and
LFs, we show the results based on equation~(\ref{eqn:sfr_guzman}) in
Figure~\ref{fig:sfr_density}. It is interesting to note how closely
this figure corresponds to Figure~\ref{fig:lum_density}.  Relative to
low redshift, the {\it AC\/}-peculiar category cause the strongest evolution even though the evolution in the {\it AC-S\/} and {\it AC-P\/} categories
is comparable at high redshift. As before, the absolute values are
highly uncertain, both because they refer only to the detected
population but also because of the approximate conversion of \oii\ to
SFR (see in particular the extensive discussion of the properties of
the CFRS objects in~\cite{CFRS-XIV}).  Nonetheless, the strong
morphological trends are clear. Finally, we note that the galaxies in
our sample satisfying the Guzman et~al.\ selection criteria contribute
26\% of the star formation rate between $z=0.5$ and $z=1.0$. This is
in reasonable agreement with Guzman {\it et.al\/}'s results when the
fact that their survey goes 1.7 magnitudes fainter than ours is taken
into account.
 
In summary, there seems to be reasonable agreement between ourselves
and Guzman et~al. in the physical properties of the faint galaxies
claimed to dominate the evolutionary trends. The overlap with our own
result is understandable, although we would argue no great significance
can be attached to the label `compact' e.g.\ in considering the present
day equivalent of such systems or in describing the sources in Figure
11.

\section{Conclusions}
\label{sec:conclusions}

We have analysed the HST images of 341 galaxies drawn from both the
CFRS and Autofib/LDSS ground-based redshift surveys. Our catalogue 
includes new spectroscopy of a magnitude-limited sample
in the Groth strip. In this, the first paper in this series, we 
have analysed the HST data, in conjunction with the available 
spectroscopic redshifts, in order to understand the evolutionary 
trends identified independently from the morphological studies in
the Medium Deep Survey and from the redshift-dependent luminosity 
functions derived from our two extensive ground-based redshift
surveys. 

We summarise our findings as follows:

\begin{itemize}
\item We have extended the automated classification scheme
developed by~\cite{Abraham-et.al-96} which places galaxies in 3 broad
categories (ellipticals/spirals/peculiars) and quantified, via
simulations and other tests, the likelihood of misclassification when
such systems are viewed at high redshift. For the typical HST exposure
times involved in this survey, systematic misclassifications occur in
the sense of shifting normal galaxies to apparently later Hubble types
at fixed {\em observed\/} wavelength. We quantify how this can be taken
into account when  redshifts are available.

\item Taking these biases into account, we demonstrate that the number
  redshift relation for regular $AC$-ellipticals is consistent with
  expectations on the basis of no evolution or `mild' evolution
  (corresponding to a magnitude of luminosity evolution by a redshift
  1).  However, the numbers are too small to differentiate these
  possibilities. In the case of the $AC$-spirals, models incorporating
  mild evolution are preferred, and the luminosity function indicate
  luminosity evolution of about 1 magnitude to $z\sim 1$.
  
\item The number of galaxies with irregular morphology increases with
  redshift well beyond what is reasonable to expect on the basis of
  systematic misclassifications of spirals. We conclude there is a
  significant evolutionary signal confined to this population.
  Analyses based on the luminosity density and luminosity functions
  divided by morphological class confirms that the demise in the {\it
    AC\/}-peculiar population is a dominant component in the recent
  evolution of the galaxy population.
  
\item There is no obvious decline with redshift in the abundance of
  regular galaxies as might be the case if the {\it AC\/}-peculiar
  population is transforming into more familiar systems. However,
  this conclusion is not particularly robust given it relies on
  uncertain volume densities in our modest sample. Such quantitative
  interpretations are also hampered by the fact that the
  classifications may well be transient and that luminosity fading may
  remove systems from the sample at lower redshift. A variety of
  different physical sources may contribute to the peculiar category
  (including misclassed late-type spirals, genuine irregulars, merging
  systems and starburst galaxies).

\end{itemize}

\acknowledgements

We acknowledge useful discussions with Simon White, Carlos Frenk, Joe
Silk and Alan Dressler . We also acknowledge the invaluable
contributions provided by all STScI staff involved in the HST
project. JB was supported by The Research Council of Norway, project
number 107798/431. Bob Abraham acknowledges support from PPARC.

\newpage
\appendix

\section{Correction of C}
\label{sec:c_correction}

A primary advantage of the use of the asymmetry and concentration
indices ($A$ and $C$) in our survey, compared to its equivalent
application to the Medium Deep Survey data, is that redshifts are
available for all the galaxies in the sample. It is therefore of 
interest to consider how to optimise the measurement of these parameters
to account for the variation of the limiting rest isophote. 

In measuring $C$, a limiting isophote, $\mu_l$, is selected relative to 
the background sky. Since the HST images generally have similar limiting
isophotes regardless of the galaxy redshift, in the rest-frame there
are shifts with redshift which scale as $10\log (1+z)$.  To correct for 
this effect, one can either attempt to measure $C$ to the same 
rest-frame isophote, or one can correct the measurements with an 
average shift calibrated with simulations. The former approach has the
disadvantage of discarding information at low redshift since, 
to maintain uniformity, the adopted rest-frame threshold must be set
fairly high to accommodate the high $z$ images. Given the large
redshift range in our sample, the latter approach is preferred. 
For symmetric light profiles the central concentration 
is given by:

\begin{equation}
  \label{eq:c_first_eqn}
  C=\frac{f(0.3R)}{f(R)},
\end{equation}
where $R$ is the radius at which the surface brightness is equal to the
limiting isophote selected, and $f(R)$ is given by:
\begin{displaymath}
  f(R)=2\pi \int_0^R I(r) r dr.
\end{displaymath}
Equation~(\ref{eq:c_first_eqn}) can be calculated exactly for a pure 
de Vaucouleurs law,
\begin{displaymath}
  I(r)=I_e\exp\left(-7.67 \left((r/r_e)^{1/4}-1\right)\right).
\end{displaymath}
The resultant expression for the central concentration is then
\begin{equation}
  \label{eq:c_for_devaucouleurs}
  C_{dV}=\frac{g(V)}{g(0.3^{1/4} V)}, \mbox{\hspace{1cm} }
  V=7.67\left(\frac{\mu_l-\mu_e}{8.3276}+1\right),
\end{equation}
with $g(V)$ being given by
\begin{displaymath}
  g(V)=\int_0^V v^7 e^{-v} dv=7! \left[ 1-e^{-x} \sum_{k=0}^7
  \frac{x^k}{k!}\right ].
\end{displaymath}

It is worth noting that the expression for $C$ is independent of
details of the profile except for the central surface brightness or,
equivalently, the surface brightness at the effective radius. This
statement remains true for exponential disks, but breaks down for
mixed profiles, where both the exponential scalelength  and the
effective radius come into play.

Accordingly, we therefore have a relation that can be used to 
calculate the change in $C$ that occurs as a function of redshift: 
$\Delta C=g(z=0; \mu_0)-g(z=z;\mu_0)$. This is a function of the 
details of the profile, but not strongly so. In practice, we used 
the functional form and calibrated the correction from the redshifted 
Frei et al images (see section~\ref{sec:redshift.biases}). This enabled
us to adopt the following relation to correct $C$:
\begin{equation}
  \label{eqn:c_correction}
  \Delta C(z)=0.6\times[C_{dV}(0;\mu_0=15.0)-C_{dV}(z;\mu_0=15.0)],
\end{equation}
with $C_{dV}$ given in equation~(\ref{eq:c_for_devaucouleurs})
above. We will refer to this as a {\em minimal \/} correction as it
does not take into account bandshifting effects, and it has been
applied to all $C$ values in this paper.
\newpage

 \begin{deluxetable}{lccccccccc}
\tablenum{1}
\scriptsize
\tablewidth{0pt}
\tablecaption{HST Imaging Survey \label{tab:hst_observations}}
\tablehead{ \colhead{Field\tablenotemark{a}} & 
  \colhead{Red filter} & 
  \colhead{Integration} &
  \colhead{\verb+NCOMB+\tablenotemark{b}} & 
  \colhead{Blue filter} & 
  \colhead{Integration} & 
  \colhead{\verb+NCOMB+\tablenotemark{b}} & 
  \colhead{$N\sub{targets}$\tablenotemark{c}} &
  \colhead{$N\sub{z}$\tablenotemark{d}} & 
  \colhead{$N\sub{star}$\tablenotemark{e}}  \\
  \colhead{} &  
  \colhead{}          &  
  \colhead{time (s)} & 
  \colhead{} &
  \colhead{} & 
  \colhead{time (s)} &
  \colhead{} &
  \colhead{} & 
  \colhead{} &
  \colhead{}} 
\startdata
cfrs\_03\_1 & F814W &  6700 & 5 & \nodata & \nodata & \nodata & 14 & 13 & 0 \nl
cfrs\_03\_2 & F814W &  6700 & 5 & \nodata & \nodata & \nodata & 15 & 14 & 0 \nl
cfrs\_03\_3 & F814W &  6700 & 5 & \nodata & \nodata & \nodata & 13 & 10 & 0 \nl
cfrs\_03\_4 & F814W &  6400 & 6 & F450W & 6600 & 6 & 11 & 11 & 0 \nl
cfrs\_03\_5 & F814W &  6400 & 6 & F450W & 6600 & 6 & 5 & 5 & 0 \nl
cfrs\_10\_1 & F814W &  6700 & 5 & \nodata & \nodata & \nodata & 17 & 16 & 1 \nl
cfrs\_10\_2 & F814W &  6700 & 5 & \nodata & \nodata & \nodata & 12 & 11 & 1 \nl
cfrs\_10\_3 & F814W &  5302.5 & 4 & \nodata & \nodata & \nodata & 24 & 20 & 2 \nl
cfrs\_14\_1 & F814W &  7400 & 6 & F450W & 7800 & 6 & 21 & 21 & 0 \nl
cfrs\_14\_2 & F814W &  7400 & 6 & F450W & 7800 & 6 & 13 & 10 & 3 \nl
cfrs\_22\_1 & F814W &  6700 & 5 & \nodata & \nodata & \nodata & 8 & 7 & 0 \nl
cfrs\_22\_2 & F814W &  6700 & 5 & \nodata & \nodata & \nodata & 14 & 12 & 0 \nl
cfrs\_22\_3 & F814W &  6700 & 5 & \nodata & \nodata & \nodata & 15 & 13 & 1 \nl
grth\_14\_1 & F814W &  4400 & 4 & F606W & 2800 & 4 & 36 & 31 & 4 \nl
grth\_14\_2 & F814W &  4400 & 4 & F606W & 2800 & 4 & 16 & 12 & 4 \nl
grth\_14\_3 & F814W &  4400 & 4 & F606W & 2800 & 4 & 21 & 17 & 1 \nl
grth\_14\_4 & F814W &  7400 & 6 & F450W & 7800 & 6 & 20 & 20 & 0 \nl
grth\_14\_5 & F814W &  4400 & 4 & F606W & 2800 & 4 & 18 & 14 & 3 \nl
grth\_14\_6 & F814W &  4400 & 4 & F606W & 2800 & 4 & 11 & 10 & 1 \nl
ldss\_10a & F814W &  5800 & 5 & F380W & 6000 & 6 & 13 & 10 & 1 \nl
ldss\_10b & F814W &  5800 & 5 & F380W & 6000 & 6 & 18 & 13 & 2 \nl
ldss\_10c & F814W &  5800 & 5 & F218W & 6000 & 6 & 10 & 7 & 2 \nl
ldss\_13a & F814W &  5800 & 5 & F380W & 6000 & 6 & 13 & 6 & 2 \nl
ldss\_13b & F814W &  5800 & 5 & F380W & 6000 & 6 & 21 & 15 & 1 \nl
ldss\_13c & F814W &  5800 & 5 & F380W & 6000 & 6 & 15 & 8 & 2 \nl
\tableline
Total\tablenotemark{f} & & & &  & & 341 & 269 & 44 & 28 \nl 
\enddata
\tablenotetext{a}{cfrs\_03=CFRS 3$^{\rm h}$ field, grth\_14=Groth
  14$^{\rm h}$ field, ldss\_10a=LDSS-2 10$^{\rm h}$.}
\tablenotetext{b}{The number of individiual exposures combined to form
  the final image}
\tablenotetext{c}{The number of spectroscopic targets in the
  field}
\tablenotetext{d}{The number of objects with reliable redshift
  measurement (note$>$1)}
\tablenotetext{e}{The number of stars}
\tablenotetext{f}{There is some overlap between the different frames
  so the total is not the sum of the columns}
\end{deluxetable}

\clearpage

  \begin{deluxetable}{rccccccccccr}
\scriptsize
\tablenum{2}

\tablecaption{Data for objects in the survey}
\tablehead{ 
\colhead{ID} & 
\colhead{z} &
\colhead{F814W} &
\colhead{M$_{B_{AB}}$} & 
\colhead{Note\tablenotemark{a}} &
\colhead{Class\tablenotemark{b}} & 
\colhead{A\tablenotemark{c}} &
\colhead{C\tablenotemark{c}} & 
\colhead{AC-Class\tablenotemark{d}} &
\colhead{EW$_{[O II]}$ (1$\sigma$ )\tablenotemark{e}} &
\colhead{Origin} &
\colhead{Old ID\tablenotemark{f}}}
\startdata

\nl
 03.0035 & 0.880 &  21.50 & -22.06 &  1 &  4 & 0.080 & 0.319 &  4 & \nodata & CFRS &    \nodata\nl
 03.0315 &  0.223 &  20.54 & -18.86 &  4 &  5 & 0.079 & 0.257 &  4 &   46(3) & CFRS &    \nodata\nl
 03.0316 &  0.815 &  22.26 & -20.91 &  3 &  4 & 0.081 & 0.202 &  4 &   12(2) & CFRS &    \nodata\nl
 03.0321 & \nodata &  21.87 & \nodata &  0 &  2 & 0.123 & 0.418 &  1 & \nodata & CFRS &    \nodata\nl
 03.0327 &  0.606 &  21.86 & -20.34 &  3 &  6 & 0.119 & 0.349 &  4 &   27(4) & CFRS &    \nodata\nl
 03.0332 &  0.188 &  21.92 & -17.62 &  4 &  4 & 0.065 & 0.216 &  4 &   14(3) & CFRS &    \nodata\nl
 03.0337 &  0.360 &  22.31 & -18.86 &  3 &  1 & 0.085 & 0.474 &  1 &   22(10) & CFRS &    \nodata\nl
 03.0346 & \nodata &  21.79 & \nodata &  0 &  4 & 0.165 & 0.418 &  4 & \nodata & CFRS &    \nodata\nl
 03.0358 & 0.088 &  17.31 & -19.95 &  3 &  2 & 0.416 & 0.550 &  1 & \nodata & CFRS &    \nodata\nl
 03.0365 &  0.219 &  19.32 & -20.08 &  4 &  3 & 0.060 & 0.449 &  1 &   32(5) & CFRS &    \nodata\nl
 03.0384 & \nodata &  21.63 & \nodata &  0 &  6 & 0.237 & 0.441 &  6 & \nodata & CFRS &    \nodata\nl
 03.0443 & 0.118 &  19.55 & -19.50 &  4 &  1 & 0.103 & 0.564 &  1 & \nodata & CFRS &    \nodata\nl
 03.0445 &  0.530 &  20.80 & -21.55 &  3 &  5 & 0.060 & 0.220 &  4 &   10(1) & CFRS &    \nodata\nl
 03.0466 &  0.534 &  23.14 & -19.59 &  3 &  4 & 0.034 & 0.198 &  4 &   29(4) & CFRS &    \nodata\nl
 03.0480 &  0.608 &  22.24 & -20.32 &  3 &  4 & \nodata & \nodata &  4 &   99(13) & CFRS &    \nodata\nl
 03.0485 &  0.606 &  21.53 & -20.24 &  4 &  6 & 0.148 & 0.146 &  6 &   78(16) & CFRS &    \nodata\nl
 03.0488 &  0.607 &  21.43 & -20.86 &  4 &  6 & 0.276 & 0.100 &  6 &   66(11) & CFRS &    \nodata\nl
 03.0523 &  0.651 &  21.28 & -21.22 &  4 &  6 & 0.313 & 0.382 &  4 &   38(10) & CFRS &    \nodata\nl
 03.0528 & 0.714 &  21.36 & -21.23 &  3 &  3 & 0.058 & 0.370 &  4 & \nodata & CFRS &    \nodata\nl
 03.0560 &  0.697 &  21.34 & -21.06 &  3 &  2 & 0.079 & 0.425 &  1 &   11(1) & CFRS &    \nodata\nl
 03.0579 &  0.660 &  22.12 & -20.47 &  2 &  4 & 0.107 & 0.236 &  4 &    5(5) & CFRS &    \nodata\nl
 03.0595 &  0.606 &  21.57 & -20.80 &  4 &  8 & 0.276 & 0.133 &  6 &   17(1) & CFRS &    \nodata\nl
 03.0599 &  0.480 &  21.19 & -20.63 &  4 &  5 & 0.111 & 0.158 &  4 &   41(5) & CFRS &    \nodata\nl
 03.0717 &  0.607 &  20.93 & -21.18 &  3 &  5 & 0.048 & 0.254 &  4 &    4(4) & CFRS &    \nodata\nl
 03.0982 &  0.195 &  21.30 & -18.12 &  4 &  2 & 0.070 & 0.429 &  1 &   34(2) & CFRS &    \nodata\nl
 03.0983 &  0.370 &  21.03 & -19.90 &  3 &  5 & 0.091 & 0.275 &  4 &    0(5) & CFRS &    \nodata\nl
 03.0992 & 0.262 &  22.74 & -17.46 &  2 &  3 & 0.149 & 0.271 &  4 & \nodata & CFRS &    \nodata\nl
 03.0999 &  0.704 &  21.49 & -21.39 &  3 &  5 & 0.308 & 0.233 &  6 &   11(2) & CFRS &    \nodata\nl
 03.1014 &  0.197 &  19.32 & -20.84 &  3 &  5 & 0.068 & 0.232 &  4 &    6(5) & CFRS &    \nodata\nl
 03.1016 &  0.705 &  22.38 & -20.47 &  3 &  6 & 0.145 & 0.161 &  6 &   85(10) & CFRS &    \nodata\nl
 03.1027 &  1.038 &  22.06 & -21.64 & 39 &  8 & 0.178 & 0.239 &  4 &  107(7) & CFRS &    \nodata\nl
 03.1031 &  0.422 &  20.62 & -20.12 &  3 &  2 & 0.105 & 0.568 &  1 &    0(12) & CFRS &    \nodata\nl
 03.1032 &  0.618 &  20.33 & -21.53 &  4 &  1 & 0.110 & 0.593 &  1 &   15(1) & CFRS &    \nodata\nl
 03.1034 & \nodata &  22.21 & \nodata &  0 &  2 & 0.057 & 0.266 &  1 & \nodata & CFRS &    \nodata\nl
 03.1035 &  0.635 &  21.25 & -20.84 &  3 &  3 & 0.095 & 0.435 &  1 &    5(2) & CFRS &    \nodata\nl
 03.1050 &  0.264 &  21.49 & -18.83 &  4 &  4 & 0.068 & 0.211 &  4 &   67(24) & CFRS &    \nodata\nl
 03.1051 &  0.155 &  21.01 & -18.16 &  4 &  5 & 0.045 & 0.272 &  4 &   29(8) & CFRS &    \nodata\nl
 03.1056 &  0.944 &  21.97 & -21.23 &  3 &  6 & 0.130 & 0.303 &  4 &   85(2) & CFRS &    \nodata\nl
 03.1060 &  0.480 &  20.72 & -20.39 &  3 &  3 & 0.095 & 0.431 &  1 &    6(4) & CFRS &    \nodata\nl
 03.1077 &  0.938 &  21.57 & -22.86 &  3 &  1 & 0.143 & 0.372 &  4 &    0(3) & CFRS &    \nodata\nl
 03.1319 &  0.620 &  21.62 & -20.61 &  4 &  1 & 0.171 & 0.425 &  1 &   32(7) & CFRS &    \nodata\nl
 03.1347 &  0.562 &  20.60 & -21.45 &  3 &  3 & \nodata & \nodata &  4 &   15(2) & CFRS &    \nodata\nl
 03.1373 &  0.482 &  20.73 & -20.42 &  3 &  3 & 0.123 & 0.536 &  1 &    0(5) & CFRS &    \nodata\nl
 03.1375 &  0.637 &  22.12 & -20.38 &  3 &  5 & 0.130 & 0.188 &  6 &   24(2) & CFRS &    \nodata\nl
 03.1381 &  0.636 &  20.23 & -22.02 &  4 &  0 & 0.074 & 0.551 &  1 &    0(5) & CFRS &    \nodata\nl
 03.1384 &  0.785 &  21.61 & -21.22 &  2 &  1 & 0.101 & 0.414 &  1 &   54(6) & CFRS &    \nodata\nl
 03.1387 & 0.222 &  20.72 & -18.60 &  1 & -1 & 0.174 & 0.622 &  1 & \nodata & CFRS &    \nodata\nl
 03.1392 &  0.605 &  20.89 & -21.22 &  4 &  4 & 0.070 & 0.389 &  1 &    0(3) & CFRS &    \nodata\nl
 03.1393 &  0.852 &  22.24 & -21.21 &  9 &  5 & 0.108 & 0.220 &  4 &    0(3) & CFRS &    \nodata\nl
 03.1395 &  0.708 &  21.79 & -20.55 &  2 &  3 & 0.092 & 0.340 &  4 &    7(0) & CFRS &    \nodata\nl
 03.1413 &  0.487 &  20.64 & -20.42 &  2 &  2 & 0.091 & 0.454 &  1 &    0(7) & CFRS &    \nodata\nl
 03.1416 &  0.488 &  20.22 & -21.20 &  4 &  1 & 0.075 & 0.484 &  1 &    2(2) & CFRS &    \nodata\nl
 03.1426 & \nodata &  22.00 & \nodata &  0 &  6 & 0.098 & 0.233 &  6 & \nodata & CFRS &    \nodata\nl
 03.1499 & 0.827 &  21.62 & -21.41 & 93 &  5 & 0.071 & 0.292 &  4 & \nodata & CFRS &    \nodata\nl
 03.1531 &  0.715 &  22.06 & -20.70 &  3 &  6 & 0.093 & 0.134 &  6 &   38(5) & CFRS &    \nodata\nl
 03.1540 &  0.690 &  21.04 & -21.56 &  3 &  4 & 0.202 & 0.286 &  4 &   18(2) & CFRS &    \nodata\nl
 03.1650 &  0.637 &  21.86 & -20.17 &  3 &  5 & 0.092 & 0.169 &  6 &   22(6) & CFRS &    \nodata\nl
 03.9003 &  0.619 &  20.88 & -21.16 &  4 &  5 & 0.283 & 0.201 &  6 &   50(2) & CFRS &    \nodata\nl
 10.0747 & 0.340 &  20.40 & -20.19 &  3 &  4 & 0.059 & 0.281 &  4 & \nodata & CFRS &    \nodata\nl
 10.0761 &  0.983 &  22.16 & -21.63 &  3 &  6 & 0.061 & 0.171 &  4 &    8(1) & CFRS &    \nodata\nl
 10.0763 &  0.671 &  21.16 & -21.85 &  3 &  5 & 0.109 & 0.232 &  4 &   11(2) & CFRS &    \nodata\nl
 10.0765 &  0.536 &  22.35 & -19.93 &  4 &  6 & \nodata & \nodata &  6 &   60(6) & CFRS &    \nodata\nl
 10.0769 &  0.669 &  21.17 & -21.28 &  4 &  2 & 0.092 & 0.306 &  4 &    4(1) & CFRS &    \nodata\nl
 10.0771 &  0.787 &  22.76 & -20.58 &  8 &  6 & 0.116 & 0.102 &  6 &   70(14) & CFRS &    \nodata\nl
 10.0793 &  0.577 &  21.31 & -20.86 &  4 &  4 & 0.196 & 0.289 &  4 &   70(9) & CFRS &    \nodata\nl
 10.0794 &  0.580 &  21.16 & -20.50 &  3 &  1 & 0.124 & 0.583 &  1 &    0(5) & CFRS &    \nodata\nl
 10.0802 &  0.309 &  21.88 & -19.28 &  4 &  6 & 0.293 & 0.147 &  6 &   47(4) & CFRS &    \nodata\nl
 10.0805 & 0.147 &  21.78 & -17.44 &  4 &  4 & 0.076 & 0.164 &  6 & \nodata & CFRS &    \nodata\nl
 10.0811 & 0.738 &  21.37 & -21.60 & 93 &  4 & 0.102 & 0.223 &  4 & \nodata & CFRS &    \nodata\nl
 10.0812 &  0.385 &  20.10 & -20.57 &  4 &  2 & 0.048 & 0.496 &  1 &    0(7) & CFRS &    \nodata\nl
 10.0813 &  0.467 &  22.27 & -19.45 &  4 &  4 & 0.118 & 0.210 &  4 &   70(14) & CFRS &    \nodata\nl
 10.0818 & 0.000 &  19.23 & \nodata &  4 & -2 & \nodata & \nodata & -2 & \nodata & CFRS &    \nodata\nl
 10.0826 &  0.643 &  20.57 & -21.95 &  3 &  5 & 0.135 & 0.232 &  4 &    7(2) & CFRS &    \nodata\nl
 10.0829 & 0.526 &  21.73 & -20.24 & 93 &  6 & 0.071 & 0.212 &  4 & \nodata & CFRS &    \nodata\nl
 10.1017 &  0.816 &  21.66 & -21.50 &  2 &  4 & 0.058 & 0.195 &  4 &    0(4) & CFRS &    \nodata\nl
 10.1153 &  0.552 &  21.03 & -20.64 &  4 &  4 & 0.099 & 0.308 &  4 &    0(3) & CFRS &    \nodata\nl
 10.1155 &  0.507 &  21.13 & -20.89 &  4 &  5 & 0.206 & 0.346 &  4 &  123(13) & CFRS &    \nodata\nl
 10.1161 &  0.200 &  20.03 & -18.91 &  4 &  2 & 0.074 & 0.457 &  1 &    5(2) & CFRS &    \nodata\nl
 10.1178 & 0.197 &  21.63 & -17.94 &  4 &  4 & 0.127 & 0.217 &  4 & \nodata & CFRS &    \nodata\nl
 10.1180 &  0.465 &  20.19 & -21.15 &  3 &  3 & 0.046 & 0.402 &  1 &    0(1) & CFRS &    \nodata\nl
 10.1182 &  0.471 &  22.26 & -19.38 &  3 &  6 & 0.086 & 0.204 &  4 &   48(18) & CFRS &    \nodata\nl
 10.1183 &  0.649 &  20.60 & -21.92 &  4 &  3 & 0.156 & 0.277 &  4 &   57(16) & CFRS &    \nodata\nl
 10.1189 &  0.949 &  21.71 & -21.98 &  2 &  3 & 0.163 & 0.351 &  4 &    0(5) & CFRS &    \nodata\nl
 10.1203 &  0.686 &  22.28 & -20.29 &  3 &  6 & 0.129 & 0.255 &  4 &   62(2) & CFRS &    \nodata\nl
 10.1207 &  0.706 &  21.42 & -20.64 &  3 &  1 & 0.155 & 0.367 &  4 &   48(16) & CFRS &    \nodata\nl
 10.1209 &  0.841 &  21.32 & -21.97 &  3 &  1 & 0.107 & 0.499 &  1 &    0(7) & CFRS &    \nodata\nl
 10.1213 &  0.817 &  21.95 & -21.04 &  3 &  6 & 0.072 & 0.175 &  4 &   27(2) & CFRS &    \nodata\nl
 10.1220 &  0.909 &  22.28 & -21.10 &  3 &  6 & 0.394 & 0.357 &  4 &   38(4) & CFRS &    \nodata\nl
 10.1222 &  0.519 &  21.42 & -20.15 &  3 &  4 & 0.180 & 0.287 &  4 &   25(16) & CFRS &    \nodata\nl
 10.1231 &  0.473 &  21.09 & -20.07 &  3 &  1 & 0.067 & 0.464 &  1 &   46(13) & CFRS &    \nodata\nl
 10.1233 & \nodata &  21.45 & \nodata &  0 &  5 & 0.078 & 0.221 &  4 & \nodata & CFRS &    \nodata\nl
 10.1236 &  0.750 &  22.12 & -20.91 &  3 &  4 & 0.215 & 0.226 &  6 &   14(4) & CFRS &    \nodata\nl
 10.1243 &  0.585 &  20.92 & -20.84 &  3 &  3 & 0.055 & 0.449 &  1 &    0(10) & CFRS &    \nodata\nl
 10.1255 &  0.467 &  19.41 & -21.95 &  4 &  0 & 0.045 & 0.475 &  1 &    9(6) & CFRS &    \nodata\nl
 10.1257 &  0.777 &  21.42 & -21.29 &  3 &  4 & 0.023 & 0.375 &  1 &    0(4) & CFRS &    \nodata\nl
 10.1262 &  0.578 &  21.65 & -20.20 &  3 &  1 & 0.116 & 0.438 &  1 &    0(5) & CFRS &    \nodata\nl
 10.1270 & 0.670 &  21.26 & -21.08 &  4 &  4 & 0.071 & 0.423 &  1 & \nodata & CFRS &    \nodata\nl
 10.1281 & 0.111 &  21.80 & -15.79 &  3 &  6 & 0.119 & 0.210 &  4 & \nodata & CFRS &    \nodata\nl
 10.1313 & \nodata &  22.37 & \nodata &  0 &  0 & 0.122 & 0.319 &  1 & \nodata & CFRS &    \nodata\nl
 10.1349 &  0.468 &  20.56 & -20.91 &  4 &  3 & 0.081 & 0.356 &  4 &   13(5) & CFRS &    \nodata\nl
 10.1423 &  0.724 &  22.59 & -20.23 &  2 &  1 & 0.100 & 0.465 &  1 &   14(3) & CFRS &    \nodata\nl
 10.1502 & \nodata\tablenotemark{g} &  22.04 & \nodata &  0 &  4 & 0.099 & 0.407 &  4 & \nodata & CFRS &    \nodata\nl
 10.1612 & 0.073 &  21.18 & -15.92 &  3 &  5 & 0.077 & 0.189 &  4 & \nodata & CFRS &    \nodata\nl
 10.1613 & 0.076 &  21.33 & -16.11 &  1 &  6 & 0.117 & 0.205 &  4 & \nodata & CFRS &    \nodata\nl
 10.1614 & 0.000 &  18.46 & \nodata &  4 & -2 & \nodata & \nodata & -2 & \nodata & CFRS &    \nodata\nl
 10.1631 & 0.000 &  20.14 & \nodata &  4 & -2 & \nodata & \nodata & -2 & \nodata & CFRS &    \nodata\nl
 10.1637 &  0.497 &  20.56 & -20.79 &  4 &  1 & 0.085 & 0.519 &  1 &    0(6) & CFRS &    \nodata\nl
 10.1643 & 0.234 &  20.61 & -18.90 &  4 &  4 & 0.146 & 0.354 &  4 & \nodata & CFRS &    \nodata\nl
 10.1644 & 0.077 &  19.90 & -17.09 &  4 &  6 & 0.066 & 0.259 &  4 & \nodata & CFRS &    \nodata\nl
 10.1650 & 0.007 &  19.21 & -14.88 &  4 &  6 & 0.353 & 0.199 &  6 & \nodata & CFRS &    \nodata\nl
 10.1651 & 0.197 &  20.28 & -18.53 &  4 &  2 & 0.093 & 0.502 &  1 & \nodata & CFRS &    \nodata\nl
 14.0147 & 1.181 &  22.26 & -21.76 &  9 &  6 & 0.122 & 0.158 &  6 & \nodata & CFRS &    \nodata\nl
 14.0163 & 0.000 &  18.99 & \nodata &  4 & -2 & \nodata & \nodata & -2 & \nodata & CFRS &    \nodata\nl
 14.0198 & 1.603 &  20.06 & \nodata & 14 & -1 & 0.159 & 0.847 & -1 & \nodata & CFRS &    \nodata\nl
 14.0207 &  0.546 &  19.67 & -22.19 &  4 &  2 & 0.056 & 0.612 &  1 &    0(15) & CFRS &    \nodata\nl
 14.0293 &  0.761 &  21.25 & -21.82 &  3 &  4 & 0.082 & 0.312 &  4 &   12(0) & GRTH &    \nodata\nl
 14.0310 &  0.238 &  20.93 & -19.40 &  4 &  6 & 0.054 & 0.236 &  4 &   36(0) & CFRS &    \nodata\nl
 14.0312 &  0.746 &  21.80 & -20.86 &  3 &  4 & 0.161 & 0.268 &  4 &   19(0) & GRTH &    \nodata\nl
 14.0377 &  0.260 &  20.81 & -19.61 &  4 &  6 & 0.218 & 0.130 &  6 &   79(30) & CFRS &    \nodata\nl
 14.0384 & 0.000 &  21.97 & \nodata &  2 & -2 & \nodata & \nodata & -2 & \nodata & CFRS &    \nodata\nl
 14.0393 &  0.602 &  20.49 & -21.93 &  4 &  5 & 0.205 & 0.147 &  4 &   22(1) & CFRS &    \nodata\nl
 14.0400 &  0.674 &  21.38 & -21.19 &  4 &  4 & 0.089 & 0.284 &  4 &   12(0) & GRTH &    \nodata\nl
 14.0411 &  0.836 &  21.45 & -21.65 &  3 &  6 & 0.212 & 0.399 &  1 &   57(0) & GRTH &    \nodata\nl
 14.0422 &  0.421 &  20.39 & -20.64 &  2 &  1 & 0.073 & 0.479 &  1 &    3(1) & CFRS &    \nodata\nl
 14.0435 & 0.068 &  18.39 & -18.83 &  3 &  3 & 0.051 & 0.454 &  1 & \nodata & CFRS &    \nodata\nl
 14.0443 & 0.000 &  20.14 & \nodata &  4 & -2 & \nodata & \nodata & -2 & \nodata & CFRS &    \nodata\nl
 14.0462 & 0.000 &  22.22 & \nodata &  3 & -2 & \nodata & \nodata & -2 & \nodata & CFRS &    \nodata\nl
 14.0485 &  0.654 &  22.20 & -20.47 &  3 &  4 & 0.037 & 0.238 &  4 &   34(5) & CFRS &    \nodata\nl
 14.0501 &  0.372 &  21.66 & -20.00 &  4 &  6 & 0.077 & 0.205 &  4 &   44(0) & GRTH &    \nodata\nl
 14.0516 & \nodata &  22.22 & \nodata &  0 &  1 & 0.116 & 0.432 &  1 & \nodata & GRTH &    \nodata\nl
 14.0528 & 0.064 &  20.61 & -16.89 &  4 &  1 & 0.107 & 0.479 &  1 & \nodata & CFRS &    \nodata\nl
 14.0529 & 0.000 &  18.45 & \nodata &  4 & -2 & \nodata & \nodata & -2 & \nodata & CFRS &    \nodata\nl
 14.0547 &  1.160 &  21.40 & -23.09 &  3 &  6 & 0.223 & 0.115 &  6 &   13(0) & GRTH &    \nodata\nl
 14.0574 & 0.000 &  21.66 & \nodata &  2 & -2 & \nodata & \nodata & -2 & \nodata & CFRS &    \nodata\nl
 14.0593 &  0.614 &  22.48 & -20.41 &  3 &  6 & 0.163 & 0.176 &  6 &   29(4) & CFRS &    \nodata\nl
 14.0608 &  0.969 &  22.16 & -21.43 &  2 &  6 & 0.118 & 0.180 &  6 &   10(0) & GRTH &    \nodata\nl
 14.0620 & 0.000 &  22.25 & \nodata &  3 & -2 & \nodata & \nodata & -2 & \nodata & CFRS &    \nodata\nl
 14.0651 & 0.637 &  22.02 & -20.30 &  1 &  2 & 0.066 & 0.437 &  1 & \nodata & CFRS &    \nodata\nl
 14.0665 &  0.809 &  22.97 & -20.78 &  2 &  6 & \nodata & \nodata &  6 &   22(0) & GRTH &    \nodata\nl
 14.0666 & 0.000 &  21.31 & \nodata &  4 & -2 & \nodata & \nodata & -2 & \nodata & CFRS &    \nodata\nl
 14.0685 & 0.081 &  17.85 & -19.96 &  4 &  2 & 0.065 & 0.524 &  1 & \nodata & CFRS &    \nodata\nl
 14.0695 &  0.266 &  21.39 & -19.08 &  4 &  1 & 0.131 & 0.347 &  4 &   40(0) & GRTH &    \nodata\nl
 14.0700 &  0.643 &  20.42 & -21.60 &  4 &  2 & 0.082 & 0.590 &  1 &    0(0) & GRTH &    \nodata\nl
 14.0725 &  0.582 &  22.11 & -19.74 &  3 &  6 & 0.190 & 0.230 &  4 &   42(3) & CFRS &    \nodata\nl
 14.0743 &  0.986 &  22.19 & -22.05 &  2 &  6 & 0.105 & 0.182 &  6 &   28(0) & GRTH &    \nodata\nl
 14.0746 &  0.675 &  21.43 & -20.68 &  3 &  2 & 0.068 & 0.423 &  1 &    0(5) & CFRS &    \nodata\nl
 14.0749 &  0.818 &  22.41 & -20.61 &  2 &  6 & 0.126 & 0.185 &  6 &   18(0) & GRTH &    \nodata\nl
 14.0760 & \nodata &  22.07 & \nodata &  0 &  8 & \nodata & \nodata &  6 & \nodata & CFRS &    \nodata\nl
 14.0807 &  0.985 &  21.88 & -21.82 &  2 &  3 & 0.088 & 0.344 &  4 &    9(0) & GRTH &    \nodata\nl
 14.0846 & 0.989 &  21.81 & -21.91 & 92 &  6 & 0.313 & 0.193 &  6 & \nodata & CFRS &    \nodata\nl
 14.0848 &  0.662 &  22.60 & -20.31 &  3 &  3 & 0.136 & 0.263 &  4 &   37(9) & CFRS &    \nodata\nl
 14.0851 & \nodata &  21.99 & \nodata &  0 &  2 & 0.121 & 0.293 &  1 & \nodata & CFRS &    \nodata\nl
 14.0854 &  0.992 &  21.63 & -22.31 &  2 &  1 & 0.096 & 0.423 &  1 &    0(5) & CFRS &    \nodata\nl
 14.0899 &  0.875 &  21.71 & -21.73 &  9 &  3 & 0.054 & 0.288 &  4 &    9(2) & CFRS &    \nodata\nl
 14.0916 &  0.325 &  20.95 & -19.94 &  3 &  3 & 0.085 & 0.328 &  4 &   18(3) & CFRS &    \nodata\nl
 14.0922 & 0.000 &  22.29 & \nodata &  4 & -2 & \nodata & \nodata & -2 & \nodata & CFRS &    \nodata\nl
 14.0939 &  0.918 &  21.96 & -22.24 &  1 &  6 & 0.268 & 0.340 &  4 &    0(0) & GRTH &    \nodata\nl
 14.0972 &  0.674 &  21.15 & -21.53 &  4 &  6 & 0.148 & 0.378 &  4 &   66(1) & CFRS &    \nodata\nl
 14.0983 &  0.286 &  21.26 & -19.59 &  4 &  3 & 0.103 & 0.373 &  4 &   34(0) & GRTH &    \nodata\nl
 14.0985 &  0.807 &  22.29 & -20.62 &  3 &  4 & 0.094 & 0.204 &  4 &   27(5) & CFRS &    \nodata\nl
 14.1008 &  0.433 &  20.66 & -20.51 &  3 &  4 & 0.090 & 0.397 &  1 &    9(0) & GRTH &    \nodata\nl
 14.1012 &  0.479 &  21.46 & -20.41 &  3 &  1 & 0.054 & 0.381 &  1 &   29(2) & CFRS &    \nodata\nl
 14.1028 &  0.988 &  21.63 & -22.22 &  3 &  2 & 0.159 & 0.415 &  1 &   31(3) & CFRS &    \nodata\nl
 14.1037 &  0.549 &  21.27 & -20.63 &  3 &  6 & 0.099 & 0.223 &  4 &   24(3) & CFRS &    \nodata\nl
 14.1039 & 0.079 &  19.57 & -18.51 &  4 &  6 & 0.153 & 0.423 &  1 & \nodata & CFRS &    \nodata\nl
 14.1042 &  0.722 &  21.33 & -21.07 &  3 &  2 & 0.137 & 0.480 &  1 &   13(1) & CFRS &    \nodata\nl
 14.1043 &  0.641 &  20.22 & -22.20 &  4 &  3 & 0.043 & 0.404 &  1 &    0(3) & CFRS &    \nodata\nl
 14.1052 & 0.000 &  17.58 & \nodata &  4 & -2 & \nodata & \nodata & -2 & \nodata & CFRS &    \nodata\nl
 14.1071 &  0.359 &  22.32 & -18.67 &  3 &  4 & 0.199 & 0.094 &  6 &   57(12) & CFRS &    \nodata\nl
 14.1079 &  0.901 &  21.82 & -21.45 &  9 &  6 & 0.086 & 0.284 &  4 &   38(3) & CFRS &    \nodata\nl
 14.1087 &  0.659 &  21.95 & -20.58 &  3 &  8 & 0.137 & 0.134 &  6 &   52(4) & CFRS &    \nodata\nl
 14.1103 &  0.209 &  22.41 & -17.75 &  4 & -1 & 0.124 & 0.564 &  1 &    0(17) & CFRS &    \nodata\nl
 14.1126 &  0.743 &  22.20 & -20.72 &  3 &  6 & 0.154 & 0.164 &  6 &   62(6) & CFRS &    \nodata\nl
 14.1129 &  0.831 &  21.12 & -22.10 &  3 &  6 & 0.251 & 0.115 &  6 &   28(0) & GRTH &    \nodata\nl
 14.1136 &  0.640 &  22.02 & -20.98 &  3 &  8 & 0.131 & 0.346 &  4 &   73(5) & CFRS &    \nodata\nl
 14.1139 &  0.660 &  20.48 & -22.31 &  3 &  6 & 0.159 & 0.227 &  4 &   16(1) & CFRS &    \nodata\nl
 14.1143 & 0.673 &  22.44 & -20.10 & 93 &  3 & 0.063 & 0.256 &  4 & \nodata & CFRS &    \nodata\nl
 14.1146 &  0.744 &  21.68 & -21.12 &  3 &  8 & 0.185 & 0.274 &  4 &   53(4) & CFRS &    \nodata\nl
 14.1158 & 0.000 &  20.81 & \nodata &  3 & -2 & \nodata & \nodata & -2 & \nodata & CFRS &    \nodata\nl
 14.1164 &  0.671 &  21.76 & -20.94 &  3 &  6 & 0.120 & 0.211 &  4 &   34(0) & GRTH &    \nodata\nl
 14.1166 &  1.015 &  22.28 & -21.36 &  3 &  1 & 0.089 & 0.358 &  4 &   56(6) & CFRS &    \nodata\nl
 14.1178 & \nodata &  22.22 & \nodata &  0 &  1 & 0.101 & 0.326 &  1 & \nodata & GRTH &    \nodata\nl
 14.1179 &  0.434 &  21.44 & -19.73 &  2 &  2 & 0.142 & 0.483 &  1 &    0(14) & CFRS &    \nodata\nl
 14.1189 &  0.753 &  22.06 & -20.80 &  3 &  4 & 0.152 & 0.302 &  4 &   43(6) & CFRS &    \nodata\nl
 14.1190 &  0.754 &  21.06 & -21.84 &  3 &  3 & 0.102 & 0.276 &  4 &    9(2) & CFRS &    \nodata\nl
 14.1193 & 0.078 &  21.67 & -16.61 &  4 &  6 & 0.220 & 0.179 &  6 & \nodata & CFRS &    \nodata\nl
 14.1200 &  0.235 &  21.84 & -18.21 &  2 &  6 & 0.096 & 0.163 &  6 &    0(42) & CFRS &    \nodata\nl
 14.1209 & 0.234 &  20.98 & -19.43 &  4 &  3 & 0.045 & 0.283 &  4 & \nodata & CFRS &    \nodata\nl
 14.1234 & 0.000 &  22.16 & \nodata &  4 & -2 & \nodata & \nodata & -2 & \nodata & CFRS &    \nodata\nl
 14.1239 &  0.362 &  21.74 & -19.68 &  3 &  6 & 0.099 & 0.220 &  4 &   47(2) & CFRS &    \nodata\nl
 14.1242 &  0.290 &  21.69 & -19.03 &  3 &  6 & 0.339 & 0.099 &  6 &   33(25) & CFRS &    \nodata\nl
 14.1251 &  0.814 &  22.17 & -20.42 &  3 &  4 & 0.076 & 0.411 &  1 &    0(0) & GRTH &    \nodata\nl
 14.1257 &  0.291 &  20.68 & -19.82 &  3 &  4 & 0.066 & 0.248 &  4 &   55(0) & GRTH &    \nodata\nl
 14.1258 &  0.645 &  22.39 & -20.25 &  3 &  1 & 0.116 & 0.323 &  4 &   62(9) & CFRS &    \nodata\nl
 14.1264 & 0.703 &  22.80 & -20.00 & 91 &  8 & \nodata & \nodata &  6 & \nodata & CFRS &    \nodata\nl
 14.1273 &  0.257 &  21.94 & -18.64 &  4 &  6 & 0.111 & 0.217 &  4 &   57(33) & CFRS &    \nodata\nl
 14.1277 &  0.810 &  21.33 & -21.93 &  2 &  8 & 0.131 & 0.387 &  1 &   19(10) & CFRS &    \nodata\nl
 14.1281 &  0.141 &  21.07 & -18.07 &  3 &  3 & 0.091 & 0.390 &  1 &   30(0) & GRTH &    \nodata\nl
 14.1311 &  0.806 &  20.58 & -22.90 &  3 &  1 & 0.087 & 0.442 &  1 &    0(1) & CFRS &    \nodata\nl
 14.1321 & 0.106 &  21.40 & -17.39 &  4 &  8 & 0.107 & 0.278 &  4 & \nodata & CFRS &    \nodata\nl
 14.1356 &  0.831 &  22.23 & -21.10 &  3 &  8 & 0.081 & 0.199 &  6 &   47(7) & CFRS &    \nodata\nl
 14.1371 & 0.000 &  18.70 & \nodata &  3 & -2 & \nodata & \nodata & -2 & \nodata & CFRS &    \nodata\nl
 14.1395 &  0.530 &  21.78 & -20.20 &  4 &  4 & 0.134 & 0.210 &  4 &   63(8) & CFRS &    \nodata\nl
 14.1415 &  0.745 &  21.06 & -21.62 &  2 &  1 & 0.138 & 0.445 &  1 &    0(0) & GRTH &    \nodata\nl
 14.1419 & 0.236 &  22.72 & -16.21 & 93 & -1 & 0.195 & 0.374 &  4 & \nodata & CFRS &    \nodata\nl
 14.1427 &  0.860 &  21.54 & -21.50 &  9 &  4 & 0.114 & 0.278 &  4 &   33(0) & GRTH &    \nodata\nl
 14.1446 &  0.348 &  20.07 & -21.10 &  4 &  1 & 0.058 & 0.468 &  1 &   24(1) & CFRS &    \nodata\nl
 14.1464 &  0.462 &  21.06 & -19.94 &  2 &  0 & 0.171 & 0.514 &  1 &   11(8) & CFRS &    \nodata\nl
 14.1496 &  0.899 &  21.93 & -21.61 &  3 & -1 & 0.105 & 0.405 &  1 &   54(6) & CFRS &    \nodata\nl
 14.1501 &  0.989 &  22.02 & -21.98 &  2 &  6 & 0.241 & 0.187 &  6 &   60(0) & GRTH &    \nodata\nl
 14.1502 & \nodata &  22.26 & \nodata &  0 &  3 & 0.087 & 0.260 &  4 & \nodata & GRTH &    \nodata\nl
 14.1524 &  0.427 &  19.87 & -21.89 &  3 &  3 & 0.124 & 0.243 &  4 &   15(0) & GRTH &    \nodata\nl
 14.9025 & 0.155 &  19.16 & -20.28 &  4 &  3 & 0.084 & 0.453 &  1 & \nodata & CFRS &    \nodata\nl
 14.9987 &  0.420 &  22.48 & -18.86 & 92 &  4 & 0.110 & 0.373 &  4 &   25(0) & CFRS &    \nodata\nl
 22.0377 & \nodata &  22.28 & \nodata &  0 &  3 & 0.077 & 0.327 &  4 & \nodata & CFRS &    \nodata\nl
 22.0434 & 0.094 &  19.90 & -18.83 &  4 &  5 & 0.062 & 0.268 &  4 & \nodata & CFRS &    \nodata\nl
 22.0453 & 0.623 &  22.12 & -20.25 &  3 &  6 & 0.150 & 0.234 &  4 & \nodata & CFRS &    \nodata\nl
 22.0497 &  0.470 &  19.42 & -22.75 &  4 &  0 & 0.171 & 0.369 &  4 &    0(5) & CFRS &    \nodata\nl
 22.0501 &  0.424 &  20.48 & -20.77 &  4 &  0 & 0.053 & 0.436 &  1 &    0(9) & CFRS &    \nodata\nl
 22.0541 & \nodata &  22.69 & \nodata &  0 &  3 & 0.099 & 0.217 &  4 & \nodata & CFRS &    \nodata\nl
 22.0576 &  0.890 &  21.93 & -21.12 &  9 &  6 & 0.225 & 0.445 &  1 &   63(25) & CFRS &    \nodata\nl
 22.0583 &  0.431 &  21.63 & -19.36 &  3 &  5 & 0.075 & 0.189 &  4 &   36(21) & CFRS &    \nodata\nl
 22.0585 &  0.294 &  20.74 & -19.40 &  2 &  2 & 0.155 & 0.348 &  4 &    0(24) & CFRS &    \nodata\nl
 22.0599 &  0.889 &  21.62 & -21.66 &  9 &  6 & 0.092 & 0.330 &  4 &   64(11) & CFRS &    \nodata\nl
 22.0609 &  0.475 &  20.60 & -20.84 &  3 &  3 & 0.085 & 0.357 &  4 &    0(11) & CFRS &    \nodata\nl
 22.0618 & 0.830 &  22.28 & -20.78 &  1 &  2 & 0.054 & 0.333 &  4 & \nodata & CFRS &    \nodata\nl
 22.0622 &  0.325 &  21.92 & -18.58 &  3 &  4 & 0.211 & 0.401 &  1 &   13(8) & CFRS &    \nodata\nl
 22.0671 &  0.319 &  20.87 & -20.24 &  4 &  3 & 0.115 & 0.371 &  4 &   31(5) & CFRS &    \nodata\nl
 22.0676 & 0.141 &  20.69 & -18.24 &  4 &  2 & 0.063 & 0.422 &  1 & \nodata & CFRS &    \nodata\nl
 22.0758 &  0.294 &  19.54 & -20.80 &  3 &  0 & 0.050 & 0.491 &  1 &    0(5) & CFRS &    \nodata\nl
 22.0764 &  0.819 &  21.98 & -20.91 &  3 &  6 & 0.128 & 0.148 &  6 &   19(3) & CFRS &    \nodata\nl
 22.0779 &  0.925 &  21.89 & -21.64 &  9 &  3 & 0.169 & 0.370 &  4 &   12(2) & CFRS &    \nodata\nl
 22.0819 &  0.293 &  20.86 & -19.61 &  4 &  4 & 0.087 & 0.239 &  4 &   45(3) & CFRS &    \nodata\nl
 22.0890 & \nodata &  21.28 & \nodata &  0 &  2 & 0.088 & 0.399 &  1 & \nodata & CFRS &    \nodata\nl
 22.0919 &  0.474 &  21.29 & -20.25 &  4 &  6 & 0.320 & 0.487 &  1 &    8(1) & CFRS &    \nodata\nl
 22.0923 & \nodata &  22.28 & \nodata &  0 &  4 & 0.149 & 0.182 &  4 & \nodata & CFRS &    \nodata\nl
 22.0944 &  0.249 &  18.84 & -21.59 &  3 &  4 & 0.061 & 0.388 &  1 &    0(57) & CFRS &    \nodata\nl
 22.0945 &  0.676 &  21.93 & -20.71 &  3 &  5 & 0.164 & 0.186 &  6 &   17(3) & CFRS &    \nodata\nl
 22.0953 &  0.977 &  22.33 & -21.39 &  8 &  6 & 0.084 & 0.154 &  6 &   34(6) & CFRS &    \nodata\nl
 22.0988 & 0.477 &  22.83 & -19.13 & 93 &  5 & 0.157 & 0.116 &  6 & \nodata & CFRS &    \nodata\nl
 22.1015 & 0.231 &  23.38 & -15.50 & 94 &  3 & \nodata & \nodata &  4 & \nodata & CFRS &    \nodata\nl
 22.1037 &  0.550 &  21.91 & -19.91 &  2 &  2 & 0.218 & 0.471 &  1 &  163(29) & CFRS &    \nodata\nl
 22.1078 & 0.671 &  22.13 & -20.65 &  1 &  1 & 0.241 & 0.477 &  1 & \nodata & CFRS &    \nodata\nl
 22.1279 &  0.594 &  21.28 & -20.70 &  3 &  1 & 0.089 & 0.358 &  1 &    0(20) & CFRS &    \nodata\nl
 22.1313 &  0.819 &  22.24 & -21.45 &  3 &  6 & 0.154 & 0.131 &  6 &   73(5) & CFRS &    \nodata\nl
 22.1374 & 0.093 &  18.21 & -20.45 &  4 &  4 & 0.054 & 0.334 &  4 & \nodata & CFRS &    \nodata\nl
 22.1406 &  0.818 &  21.97 & -20.97 &  4 & -1 & 0.121 & 0.319 &  4 &  100(4) & CFRS &    \nodata\nl
 22.1453 &  0.816 &  21.61 & -21.59 &  3 &  6 & 0.340 & 0.163 &  6 &    0(3) & CFRS &    \nodata\nl
 22.1466 & \nodata &  21.85 & \nodata &  0 &  3 & 0.135 & 0.273 &  4 & \nodata & CFRS &    \nodata\nl
 22.1486 &  0.953 &  22.58 & -21.32 &  8 & -1 & \nodata & \nodata & -1 &   12(6) & CFRS &    \nodata\nl
 22.1507 &  0.820 &  21.37 & -21.48 &  3 &  1 & 0.098 & 0.446 &  1 &    0(2) & CFRS &    \nodata\nl
10.10116 & \nodata &  19.75 & \nodata &  0 &  4 & 0.085 & 0.332 &  4 & \nodata & Autofib &    10f\_14\nl
10.11699 & 0.000 &  19.76 & \nodata &  4 & -2 & \nodata & \nodata & -2 & \nodata & LDSS1 &   10.2.9HI\nl
10.11702 &  0.168 &  19.36 & \nodata &  4 &  2 & 0.118 & 0.511 &  1 &   10(-9) & LDSS1 &  10.2.12HI\nl
10.11703 &  0.437 &  19.63 & \nodata &  4 &  3 & 0.050 & 0.429 &  1 &    5(-9) & LDSS1 &  10.2.13HI\nl
10.11706 &  0.151 &  21.27 & \nodata &  4 &  6 & 0.230 & 0.133 &  6 &   21(-9) & LDSS1 &  10.2.16HI\nl
10.11709 &  0.179 &  20.94 & \nodata &  4 &  3 & 0.046 & 0.276 &  4 &   16(-9) & LDSS1 &  10.2.19HI\nl
10.12058 & \nodata &  23.36 & \nodata &  0 &  6 & 0.268 & 0.081 &  6 & \nodata & LDSS2 &  10.21.227\nl
10.12059 &  0.307 &  21.15 & \nodata &  4 &  5 & 0.049 & 0.218 &  4 &   20(2) & LDSS2 &  10.21.233\nl
10.12060 &  0.294 &  21.23 & -18.39 &  4 &  1 & 0.120 & 0.493 &  1 &   12(4) & LDSS2 &  10.21.262\nl
10.12062 &  0.634 &  21.85 & -20.21 &  4 &  0 & 0.110 & 0.475 &  1 &   55(1) & LDSS2 &  10.21.279\nl
10.12063 &  1.108 &  22.27 & -21.36 &  4 & -1 & 0.161 & 0.397 &  1 &   65(10) & LDSS2 &  10.21.288\nl
10.12065 &  0.207 &  20.97 & -17.85 &  4 &  1 & 0.103 & 0.587 &  1 &   19(6) & LDSS2 &  10.21.328\nl
10.12066 &  0.924 &  22.30 & -20.36 &  2 &  6 & 0.104 & 0.155 &  6 &   17(8) & LDSS2 &   10.21.22\nl
10.12071 &  0.177 &  21.25 & -17.84 &  4 &  3 & 0.082 & 0.283 &  4 &    0(8) & LDSS2 &   10.21.88\nl
10.12073 &  0.492 &  20.35 & -20.61 &  4 &  6 & 0.074 & 0.354 &  4 &    0(0) & LDSS2 &  10.21.109\nl
10.12076 &  0.323 &  21.91 & -18.40 &  4 &  5 & 0.067 & 0.208 &  4 &   25(5) & LDSS2 &  10.21.301\nl
10.12078 &  0.296 &  22.82 & \nodata &  4 &  3 & 0.072 & 0.196 &  4 &   58(8) & LDSS2 &  10.22.223\nl
10.12080 &  0.314 &  20.81 & -18.96 &  4 & -1 & 0.182 & 0.643 &  1 &    0(3) & LDSS2 &  10.22.248\nl
10.12081 &  0.563 &  22.23 & -19.42 &  4 &  6 & 0.221 & 0.142 &  6 &   17(2) & LDSS2 &  10.22.260\nl
10.12085 & \nodata &  23.02 & \nodata &  0 &  6 & 0.259 & 0.109 &  6 & \nodata & LDSS2 &  10.22.315\nl
10.12086 &  0.324 &  22.13 & -18.24 &  4 &  6 & 0.105 & 0.159 &  6 &   50(3) & LDSS2 &  10.22.330\nl
10.12087 & 2.749 &  22.81 & \nodata &  4 & -1 & \nodata & \nodata & -1 & \nodata & LDSS2 &   10.22.25\nl
10.12089 &  0.384 &  20.39 & -19.99 &  4 &  1 & 0.060 & 0.500 &  1 &    0(3) & LDSS2 &   10.22.61\nl
10.12091 &  0.476 &  20.25 & -20.76 &  4 &  3 & 0.104 & 0.456 &  1 &   10(2) & LDSS2 &   10.22.71\nl
10.12092 &  0.436 &  20.23 & \nodata &  4 &  2 & 0.090 & 0.490 &  1 &    0(2) & LDSS2 &   10.22.77\nl
10.12095 &  0.724 &  21.75 & -20.59 &  2 &  6 & 0.073 & 0.218 &  4 &   30(2) & LDSS2 &  10.22.122\nl
10.12519 &  0.097 &  19.74 & -18.45\tablenotemark{h} &  4 &  3 & 0.182 & 0.315 &  4 &    0(0) & LDSS2 &  10.23.218\nl
10.12520 & 1.999 &  22.42 & \nodata &  4 & -1 & \nodata & \nodata & -1 & \nodata & LDSS2 &  10.23.222\nl
10.12522 & 0.000 &  19.02 & \nodata &  4 & -2 & \nodata & \nodata & -2 & \nodata & LDSS2 &  10.23.235\nl
10.12524 &  0.149 &  21.51 & -17.16 &  4 &  2 & 0.108 & 0.326 &  4 &   42(5) & LDSS2 &  10.23.255\nl
10.12525 &  0.435 &  19.79 & -20.81 &  4 &  1 & 0.291 & 0.069 &  1 &    3(1) & LDSS2 &  10.23.273\nl
10.12527 & 0.000 &  19.87 & \nodata &  4 & -2 & \nodata & \nodata & -2 & \nodata & LDSS2 &  10.23.332\nl
10.12528 &  0.582 &  20.69 & -21.11 &  4 &  5 & 0.088 & 0.317 &  4 &    8(1) & LDSS2 &   10.23.28\nl
10.12529 & \nodata &  23.16 & \nodata &  0 &  6 & 0.243 & 0.163 &  6 & \nodata & LDSS2 &   10.23.32\nl
10.12530 &  0.476 &  20.50 & -20.42 &  4 &  3 & 0.123 & 0.468 &  1 &    0(0) & LDSS2 &   10.23.40\nl
10.12534 & \nodata &  21.31 & \nodata &  0 & -1 & 0.238 & 0.549 & -1 & \nodata & LDSS2 &   10.23.92\nl
10.12535 & 0.000 &  18.07 & \nodata &  4 & -2 & \nodata & \nodata & -2 & \nodata & LDSS2 &  10.23.105\nl
10.12536 & 1.256 &  21.00 & \nodata &  4 & -1 & 0.158 & 0.622 &  1 & \nodata & LDSS2 &  10.23.116\nl
10.12537 & 0.000 &  21.69 & \nodata &  4 & -2 & \nodata & \nodata & -1 & \nodata & LDSS2 &  10.23.126\nl
10.12786 & \nodata &  21.88 & \nodata &  0 &  6 & \nodata & \nodata &  6 & \nodata & LDSS1 &   10.2.2FB\nl
10.12787 & 0.283 &  22.32 & \nodata &  4 & -1 & \nodata & \nodata & -1 & \nodata & LDSS1 &   10.2.5FB\nl
13.10222 &  0.052 &  18.98 & -17.23\tablenotemark{i} & -1 &  3 & 0.059 & 0.249 &  4 &    0(-9) & Autofib &    13b\_14\nl
13.10379 & \nodata &  19.35 & \nodata &  0 & -2 & \nodata & \nodata & -2 & \nodata & Autofib &    13m\_16\nl
13.11753 &  0.198 &  19.08 & \nodata &  2 &  4 & \nodata & \nodata &  4 &   28(-9) & LDSS1 &   13.2.1HI\nl
13.11772 &  0.512 &  20.90 & \nodata &  4 &  4 & 0.098 & 0.353 &  4 &   63(-9) & LDSS1 &  13.2.20HI\nl
13.11874 & \nodata &  19.33 & \nodata &  0 &  3 & 0.064 & 0.415 &  4 & \nodata & Autofib &    13f1\_1\nl
13.11924 &  0.281 &  18.86 & -19.83\tablenotemark{i} & -1 &  1 & 0.085 & 0.634 &  1 &    0(-9) & Autofib &   13f1\_63\nl
13.11925 &  0.256 &  20.34 & -18.83\tablenotemark{i} & -1 &  3 & 0.061 & 0.320 &  4 &   46(-9) & Autofib &   13f1\_64\nl
13.11976 &  0.336 &  21.37 & -18.93\tablenotemark{i} & -1 &  5 & 0.090 & 0.247 &  4 &  100(-9) & Autofib &   13xf\_64\nl
13.12099 &  0.385 &  21.00 & -19.69 &  4 &  5 & 0.095 & 0.345 &  4 &   19(2) & LDSS2 &  13.21.323\nl
13.12106 &  0.556 &  21.60 & -20.09 &  4 &  6 & 0.103 & 0.200 &  4 &   14(3) & LDSS2 &  13.21.465\nl
13.12107 &  0.556 &  21.22 & -20.47 &  2 &  2 & 0.143 & 0.389 &  1 &   34(5) & LDSS2 &  13.21.480\nl
13.12109 &  0.462 &  22.25 & -18.97 &  2 &  5 & 0.153 & 0.143 &  4 &   36(7) & LDSS2 &  13.21.517\nl
13.12111 &  0.089 &  22.31 & -15.42 &  4 &  6 & 0.191 & 0.154 &  6 &   41(7) & LDSS2 &   13.21.27\nl
13.12112 &  0.424 &  21.09 & -19.92 &  4 &  2 & 0.102 & 0.474 &  1 &   12(2) & LDSS2 &   13.21.38\nl
13.12116 &  0.187 &  21.18 & -17.94 &  4 &  1 & 0.058 & 0.472 &  1 &   59(14) & LDSS2 &  13.21.106\nl
13.12117 &  0.536 &  22.05 & -19.55 &  2 &  5 & 0.094 & 0.224 &  4 &   32(3) & LDSS2 &  13.21.123\nl
13.12118 &  0.335 &  21.86 & -18.61 &  4 &  5 & 0.135 & 0.333 &  4 &   35(4) & LDSS2 &  13.21.160\nl
13.12538 & \nodata &  21.71 & \nodata &  0 &  2 & 0.085 & 0.490 &  1 & \nodata & LDSS2 &  13.21.311\nl
13.12539 & 0.000 &  19.97 & \nodata &  4 & -2 & \nodata & \nodata & -2 & \nodata & LDSS2 &  13.22.325\nl
13.12540 &  0.452 &  22.33 & -18.99 &  2 &  6 & 0.148 & 0.178 &  6 &    0(0) & LDSS2 &  13.22.344\nl
13.12542 & \nodata &  22.03 & \nodata &  0 &  6 & 0.200 & 0.153 &  6 & \nodata & LDSS2 &  13.22.367\nl
13.12545 &  0.830 &  20.14 & -22.57 &  4 &  6 & 0.195 & 0.289 &  4 &   19(1) & LDSS2 &  13.22.400\nl
13.12546 &  0.283 &  20.01 & -19.44 &  4 &  3 & 0.112 & 0.528 &  1 &    0(0) & LDSS2 &  13.22.417\nl
13.12549 &  0.493 &  21.02 & -20.26 &  4 &  3 & 0.105 & 0.328 &  4 &   13(1) & LDSS2 &  13.22.469\nl
13.12550 & 0.000 &  18.44 & \nodata &  4 & -2 & \nodata & \nodata & -2 & \nodata & LDSS2 &  13.22.484\nl
13.12551 & \nodata &  22.12 & \nodata &  0 &  5 & 0.077 & 0.224 &  4 & \nodata & LDSS2 &  13.22.492\nl
13.12552 &  0.566 &  20.42 & -21.14 &  4 &  3 & 0.093 & 0.268 &  4 &    7(1) & LDSS2 &  13.22.510\nl
13.12553 &  0.278 &  20.47 & -19.42 &  4 &  3 & 0.086 & 0.310 &  4 &    8(3) & LDSS2 &  13.22.519\nl
13.12554 & \nodata &  20.90 & \nodata &  0 &  4 & 0.318 & 0.233 &  4 & \nodata & LDSS2 &   13.22.12\nl
13.12555 &  0.426 &  21.99 & -19.09 &  2 & -1 & 0.165 & 0.385 &  1 &    4(0) & LDSS2 &   13.22.28\nl
13.12556 & \nodata &  22.57 & \nodata &  0 & -2 & \nodata & \nodata & -2 & \nodata & LDSS2 &   13.22.34\nl
13.12559 & \nodata &  22.73 & \nodata &  0 & -1 & \nodata & \nodata & -1 & \nodata & LDSS2 &   13.22.98\nl
13.12560 &  0.363 &  20.49 & -19.90 &  4 &  1 & 0.067 & 0.528 &  1 &    7(2) & LDSS2 &  13.22.116\nl
13.12561 &  0.326 &  19.51 & -20.73 &  4 &  3 & 0.106 & 0.519 &  1 &    4(1) & LDSS2 &  13.22.131\nl
13.12566 & 0.000 &  19.59 & \nodata &  4 & -2 & \nodata & \nodata & -2 & \nodata & LDSS2 &  13.22.180\nl
13.12567 & \nodata &  22.28 & \nodata &  0 &  1 & 0.152 & 0.422 &  1 & \nodata & LDSS2 &  13.22.186\nl
13.12759 & 0.000 &  21.07 & \nodata &  4 & -2 & \nodata & \nodata & -2 & \nodata & LDSS1 &   13.2.9LO\nl
13.12764 & \nodata &  21.97 & \nodata &  0 &  2 & 0.156 & 0.457 &  1 & \nodata & LDSS1 &  13.2.14LO\nl
13.12767 & \nodata &  21.84 & \nodata &  0 & -1 & \nodata & \nodata & -1 & \nodata & LDSS1 &  13.2.17LO\nl
13.12783 & \nodata &  22.45 & \nodata &  0 &  8 & 0.210 & 0.205 &  6 & \nodata & LDSS1 &  13.2.35LO\nl
13.12795 & 2.934 &  21.85 & \nodata &  4 & -1 & \nodata & \nodata & -1 & \nodata & LDSS1 &   13.2.1FB\nl
13.12797 & 0.627 &  22.69 & \nodata &  2 & -1 & \nodata & \nodata & -1 & \nodata & LDSS1 &   13.2.3FB\nl
13.12798 & 0.297 &  22.78 & \nodata &  2 & -1 & \nodata & \nodata & -1 & \nodata & LDSS1 &   13.2.4FB\nl
13.12801 & \nodata &  22.47 & \nodata &  0 &  6 & 0.157 & 0.167 &  6 & \nodata & LDSS1 &   13.2.7FB\nl
13.12802 & 0.667 &  22.63 & \nodata &  2 &  2 & 0.120 & 0.233 &  4 & \nodata & LDSS1 &   13.2.9FB\nl
13.12803 & \nodata &  23.10 & \nodata &  0 &  6 & \nodata & \nodata &  6 & \nodata & LDSS1 &  13.2.11FB\nl
13.12808 & 0.000 &  22.18 & \nodata &  0 & -2 & \nodata & \nodata & -2 & \nodata & LDSS1 &  13.2.28LO\nl
13.12810 & \nodata &  22.30 & \nodata &  0 & -1 & \nodata & \nodata & -1 & \nodata & LDSS1 &  13.2.10FB\nl
13.12811 & 0.550 &  21.90 & \nodata &  2 &  4 & 0.137 & 0.193 &  4 & \nodata & LDSS1 &  13.2.13FB   \enddata 
  \tablenotetext{a}{The confidence class for the redshift. For the LDSS 
                    objects this has been transformed to the CFRS system
                    by assigning note=4 to confident redshifts, note=2 to 
                    less secure redshifts and 0 for failures. For the few
                    LDSS objects for which there is no confidence class, we 
                    have assigned note=-1}
  \tablenotetext{b}{The eyeball classification for the object}
  \tablenotetext{c}{The A\&C parameters (uncorrected), see
                    Section~3.2}
  \tablenotetext{d}{The AC classification for the object using the division 
                    lines in Figure~8}
  \tablenotetext{e}{The equivalent width of \oii. For the CFRS objects this 
                    is from Hammer et~al. (1997), for the LDSS objects
                     from the Autofib survey (Ellis et~al. 1996). }
  \tablenotetext{f}{The identification given in the original LDSS paper}
  \tablenotetext{g}{The object is clearly extended, but was given $z=0$ in
                    in the CFRS survey.}
  \tablenotetext{h}{The HST photometry here is uncertain and $M_{B_{AB}}$ is 
                     based on the original $b_J$ photometry.}
  \tablenotetext{i}{The absolute magnitude is the original Autofib absolute
                     magnitude based on $b_J$ transformed to AB.}
   \end{deluxetable}
\clearpage

\begin{deluxetable}{lcccc}
\scriptsize
\tablenum{3}
\tablecaption{HST Survey Completeness}
\tablehead{\colhead{Survey} & \colhead{Survey limits} &
  \multicolumn{2}{c}{Completeness} & \colhead{Effective
    area\tablenotemark{a}} \\ 
\colhead{}  & \colhead{} & \colhead{Geometric} &
\colhead{Spectroscopic}   & \colhead{deg$^2$}}
\startdata
  CFRS & $17.5<I_{AB}<22.5$ & 0.5477 & 90.44\% & 0.01377 \nl
  LDSS-2 (10hr) & $22.5<b_J<24.0$ & 0.7941 & 85.19\% & 0.002513 \nl
  LDSS-2 (13hr)\tablenotemark{b} & $22.5<b_J<23.3$ & 0.4516 & 85.71\% & 0.001576\nl
\enddata
\tablenotetext{a}{Defined as surveyed area times the geometric completeness}
\tablenotetext{b}{The LDSS-2 13hr field have less deep spectroscopy,
  we have adopted the completeness limits as discussed in
  Glazebrook et~al. (1995b)}
\end{deluxetable}

\clearpage

\begin{deluxetable}{cccccccc}
\tablenum{4}
\tablecaption{The movement of Frei galaxies in the AC plane
\label{tab:redshifting_results} }
\scriptsize
\tablewidth{\textwidth}
\tablehead{\colhead{ } &
\colhead{AC-E} &
\colhead{AC-S} &
\colhead{AC-P} &
\colhead{$\cf{SP}^{\rm b}$} &
\colhead{$\cf{PS}^{\rm b}$} &
\colhead{$\cf{SE}^{\rm b}$} &
\colhead{$\cf{ES}^{\rm b}$}}
\startdata
\multicolumn{8}{c}{Drift coefficients for corrections to $R$ rest
  frame morphologies} \nl
z=0.0\tablenotemark{a} & 33 & 41 & 5 & $0 $ & $0 $ & $0 $ & 
$0 $ \nl 
z=0.2 & 29 & 30 & 6 & $0 $ & $0 $ & $0.06 \pm 0.04 $ & 
$0 $ \nl 
z=0.7 & 30 & 29 & 5 & $0.13 \pm 0.09 $ & $0 $ & $0.20 \pm 0.12 $ & 
$0.32 \pm 0.13 $ \nl 
z=0.9 & 16 & 22 & 6 & $0.24 \pm 0.11 $ & $0 $ & $0.10 \pm 0.07 $ & 
$0.33 \pm 0.12 $ \nl 
\multicolumn{8}{c}{Drift coefficients for corrections to $B_J$ rest
  frame morphologies} \nl
z=0.0\tablenotemark{a} & 33 & 41 & 5 & $0 $ & $0.67 \pm 0.21 $ & $0.29 \pm 0.08 $ & 
$0.05 \pm 0.05 $ \nl 
z=0.2 & 29 & 30 & 6 & $0 $ & $0.67 \pm 0.24 $ & $0.35 \pm 0.10 $ & 
$0 $ \nl 
z=0.7 & 30 & 29 & 5 & $0.05 \pm 0.05 $ & $0.67 \pm 0.47 $ & $0.43 \pm 0.14 $ & 
$0.30 \pm 0.17 $ \nl 
z=0.9 & 16 & 22 & 6 & $0.07 \pm 0.05 $ & $0.20 \pm 0.20 $ & $0.26 \pm 0.10 $ & 
$0.21 \pm 0.12 $ \nl 
\enddata
\tablenotetext{a}{The numbers for $z=0$ are for all Frei galaxies.}
\tablenotetext{b}{The errors are 1$\sigma$ Poisson errors.}
\end{deluxetable}

\clearpage

\begin{deluxetable}{lcccc}
\tablenum{5}
\scriptsize
\tablecaption{Adopted local LF ($H_0=50$km/s/Mpc)\label{tab:lf_params}}
\tablehead{ \colhead{Hubble type} & 
\colhead{$M^\star$ ($b_J$)} & 
\colhead{$\alpha$} & 
\colhead{$\phi^\star (Mpc^{-3})$}} 
\startdata
E/S0 & $-21.21$ & $-1.00$ & $1.39\times 10^{-3}$ \nl
Sab & $-20.90$ & $-1.00$ &$6.0\times 10^{-4}$ \nl
Sbc & $-20.90$ &$-1.00$ & $1.1\times 10^{-3}$ \nl
Scd & $-20.90$ &$-1.00$ & $4.5\times 10^{-4}$ \nl
Sdm & $-20.90$ &$-1.00$ & $2.6\times 10^{-4}$ \nl
Irr & $-20.29$ &$-1.87$ & $7.5\times 10^{-5}$ \nl
\enddata
\end{deluxetable}

\clearpage

\figcaption{The absolute magnitude - redshift distribution for the
  combined CFRS and LDSS-2 surveys (assuming $H_o$=50 kms s$^{-1}$
  Mpc$^{-1}$).  Open symbols refer to the CFRS sample limited at
  I(AB)=22.5; filled symbols refer to the LDSS-2 survey limited at
  $b_J$=24. The objects in the HST survey are indicated with circles.
  \label{fig:ldss_vs_cfrs_redshift} } 

\figcaption{The relative contributions of the LDSS-2 and CFRS redshift
  surveys to the total redshift distribution for the sample used in
  the analysis.
  \label{fig:redshift_contribution} }

\figcaption{Verifying the absolute magnitude scales of the CFRS and
  LDSS-2 redshift surveys. Top panel: $B_{AB}-b_J$ colour for the
  LDSS-2 galaxies. Middle panel: Difference in $M_B$ for CFRS galaxies
  using the CFRS and LDSS-2 SEDs. Bottom panel: The redshift
  dependence of $M_B$ differences obtained by calculating
  $k$-corrections for the LDSS-2 galaxies using $b_J$ and $I_{F814W}$.
  \label{fig:magnitude_transforms} } 

\figcaption{Examples of the morphological types used in this paper,
  with the total number of objects in each class in the survey. Each
  image is $6\times6$ arcsec$^2$.
  \label{fig:examples_of_morphology} } 

\figcaption{Correlation between the different classifiers. The radii
  of the circles are proportional to the value in the correlation
  matrix. The correlation between the different classifiers is only
  slightly larger than the internal scatter of one classifier.
  \label{fig:correlations} }

\figcaption{The distribution of galaxies in the $M(B_{AB})-z$ plane
  for the HST survey. Inlaid histograms show the fractional
  contribution of the four different visual classes as a function of
  redshift. In this plot the tadpoles have been grouped with the
  irregular galaxies.  \label{fig:mbz_figure} }

\figcaption{Number magnitude counts for the survey (filled circles)
  compared with those from the larger MDS survey (open triangles).
  Error bars are Poissonian (\protect\cite{Gehrels-86}). In this
  plot the tadpoles have been grouped with the irregulars.
  \label{fig:nm_variation} } 

\figcaption{Upper panel: The distribution of galaxies in the
  $A-C$ plane. Only objects with area larger than 64 contiguous pixels
  are shown. Bottom panel: The classification angle $\Theta$
  (see text) plotted versus the eyeball classification for each
  object. The large circles show the median of each eyeball class.
  \label{fig:ac_plot} }

\figcaption{The absolute magnitude distribution of the Frei {\it et
  al\/} 
  galaxies with the selection limits appropriate for the deep HST
  survey overlaid.  The selection function for Sbc galaxies is
  indicated.
  \label{fig:bobs_figure} }

\figcaption{Top panel: The change in asymmetry $A$ between the $R$ and
  $B_J$ images of the local Frei {\it et al\/} galaxies. Filled circles denote
  late type systems for which the trends are somewhat more pronounced.
  Bottom Panel: as above for the concentration $C$ illustrating the
  effect of the bandshifting effects on A\&C classifications.
  Labelled lines define the boundary where the change in $C$ would
  move an object to a different class as indicated (assuming $A$ is at
  the median of the distribution)
  \label{fig:deltaa_deltac} } 

\figcaption{The objects classed as {\it AC\/}-peculiar sorted by
  increasing redshift upwards the page. The left column contains
  objects with EW\oii$=0$ or unknown. The galaxies in the middle
  column have $0<$EW\oii$<40$ and the right galaxies the objects with
  EW\oii$>40$.\label{fig:ac_irregulars}. Every image is $6\times6$
  arcsec$^2$, with the exception of 10.1650 which is $12\times 12$
  arcsec$^2$.}

\figcaption{Redshift distribution of the three broad $AC$ classes
  compared with theoretical predictions for no-evolution (dashed line)
  and for mild evolution corresponding to one magnitude of luminosity
  evolution at $z=1$ (dash-dotted line).  The models have been
  corrected to observed numbers using the method outlined in the text.
  \label{fig:nz_distributions} }

\figcaption{Luminosity functions as a function of $AC$ class binned as
  in the CFRS analysis (\protect\cite{CFRS-VI}). The dotted line
  is the CFRS $0.2<z<0.5$ luminosity function.
  \label{fig:luminosity_functions} }

\figcaption{The $B_{AB}$ rest-frame luminosity density of galaxies
  detected in the survey as a function of redshift and $AC$ class.
  The values plotted are offset in redshift slightly for clarity. The
  downward arrow indicates the effect of a bandshifting correction of
  24\% for the $AC-P$ class The open symbols represent the corrected
  luminosity densities, as discussed in the text.
  \label{fig:lum_density} }

\figcaption{Star formation density for detected sources as a function
  of redshift for the three $AC$ classes calculated using
  equation~(\protect\ref{eqn:sfr_guzman}). The upward arrows show the
  total in each redshift bin. The downward arrow indicates a change of
  24\% in $\rho_{\rm SFR}$ for $AC$-peculiars \label{fig:sfr_density}
  }

\psfig{file=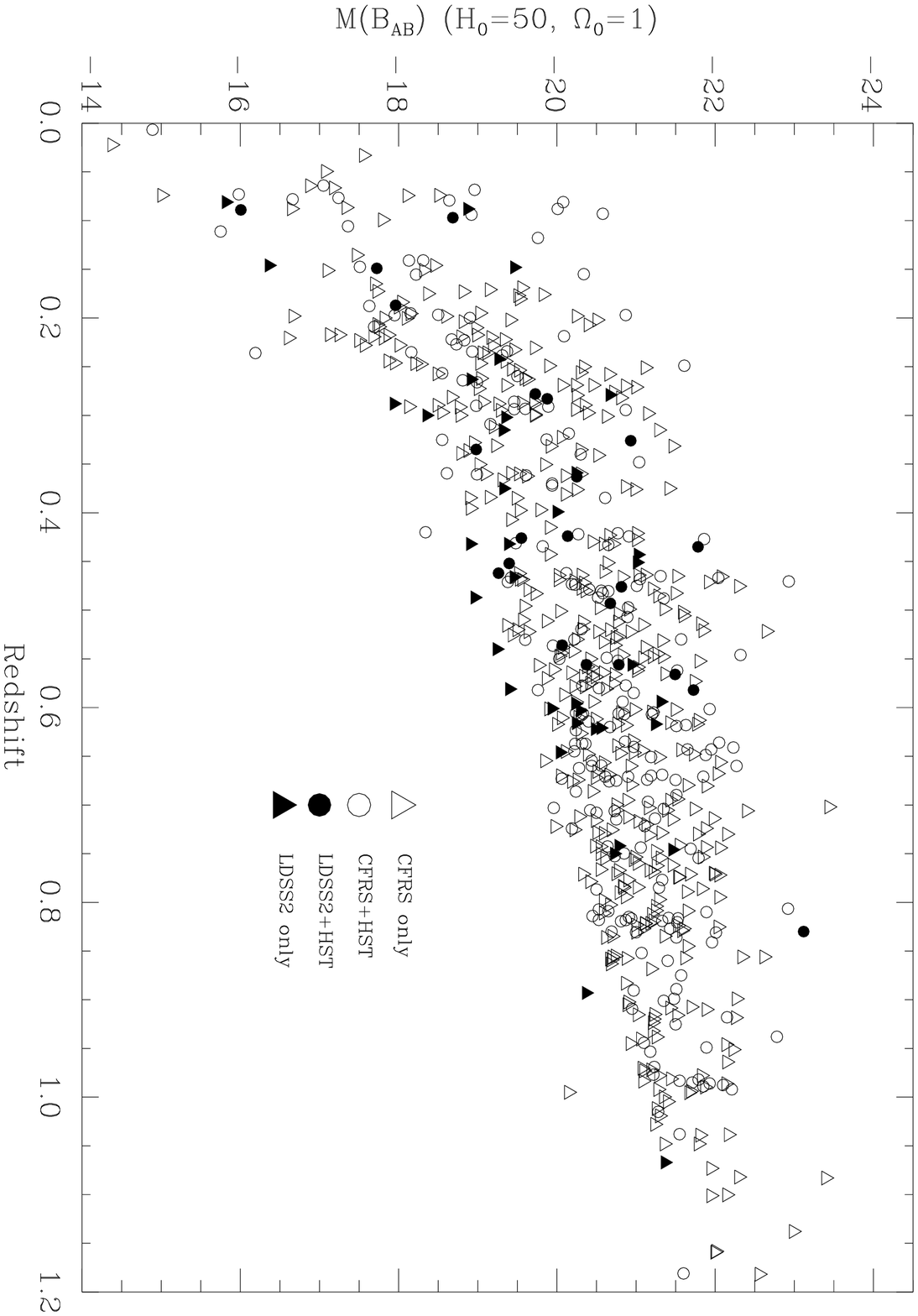,height=0.9\textheight}
\psfig{file=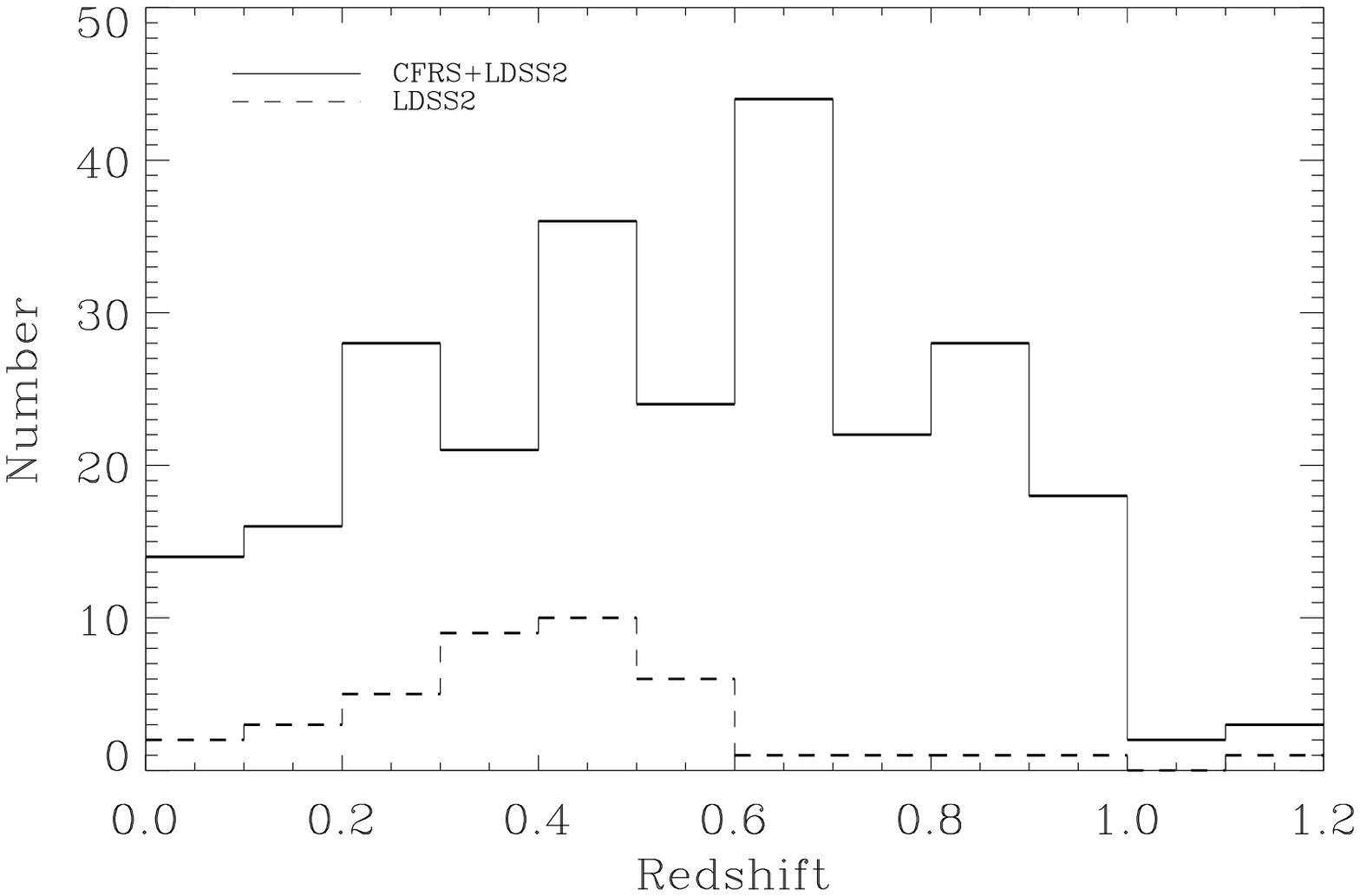,width=0.9\textwidth}
\psfig{file=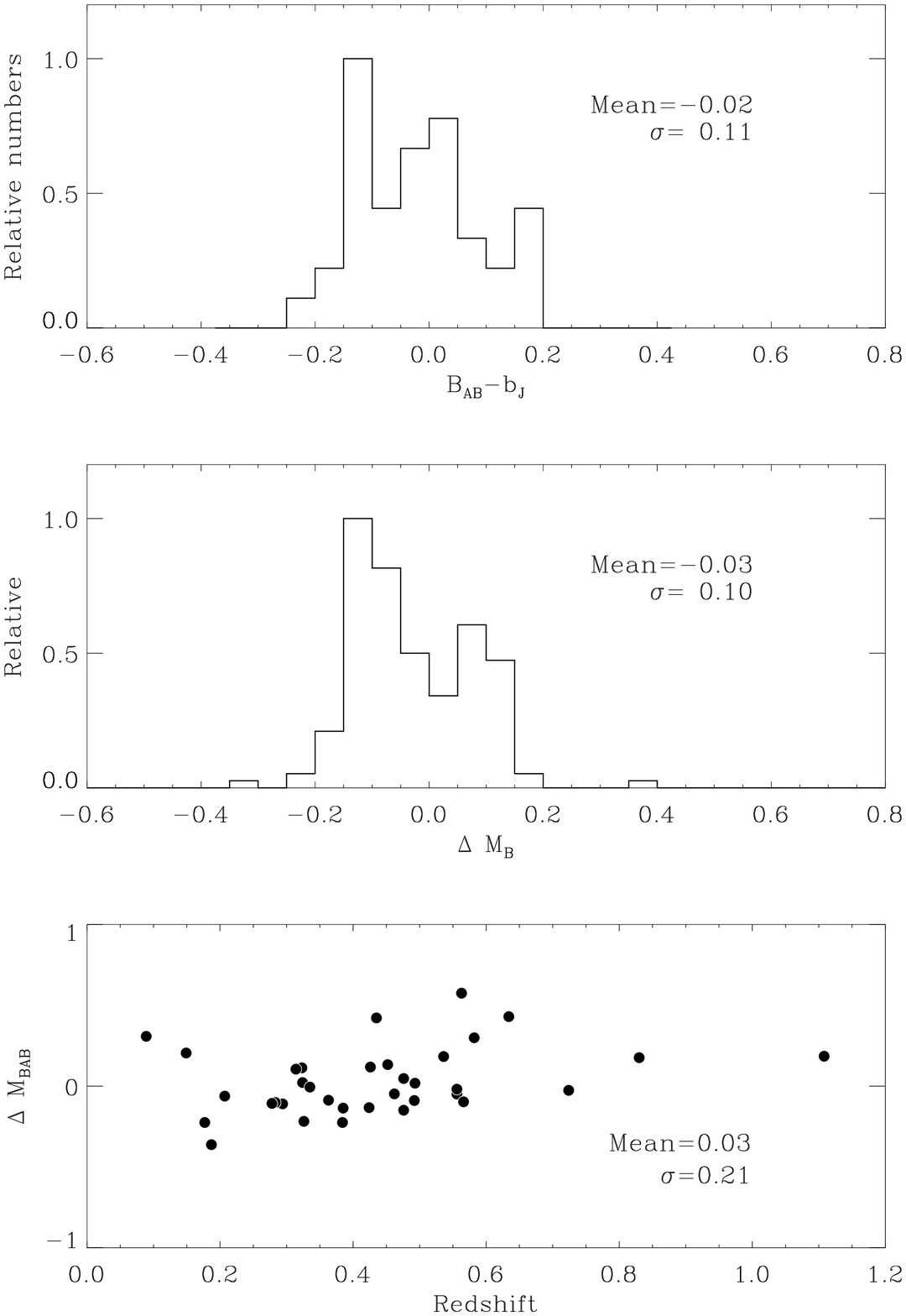,height=0.9\textheight}
\psfig{file=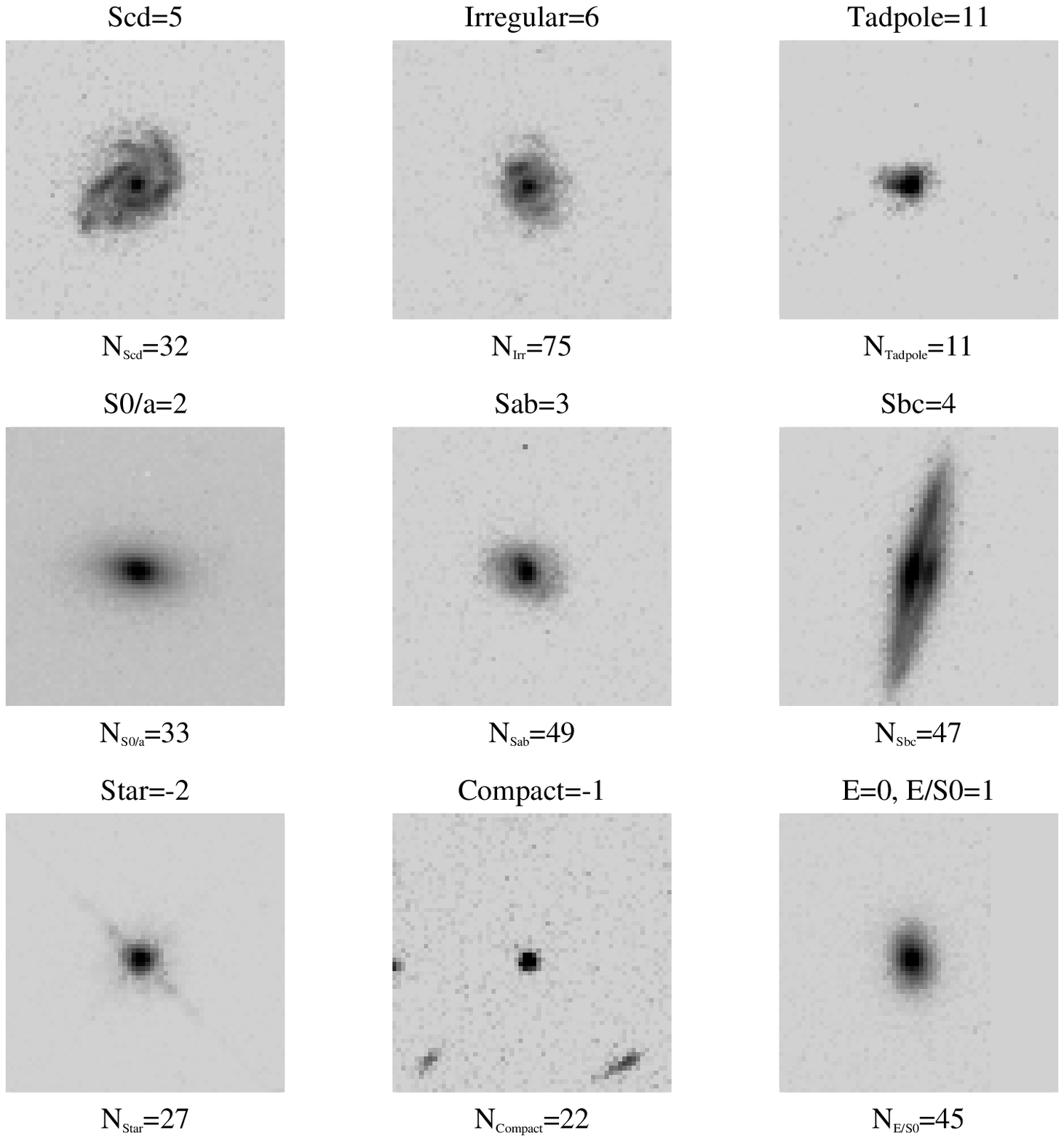}
\psfig{file=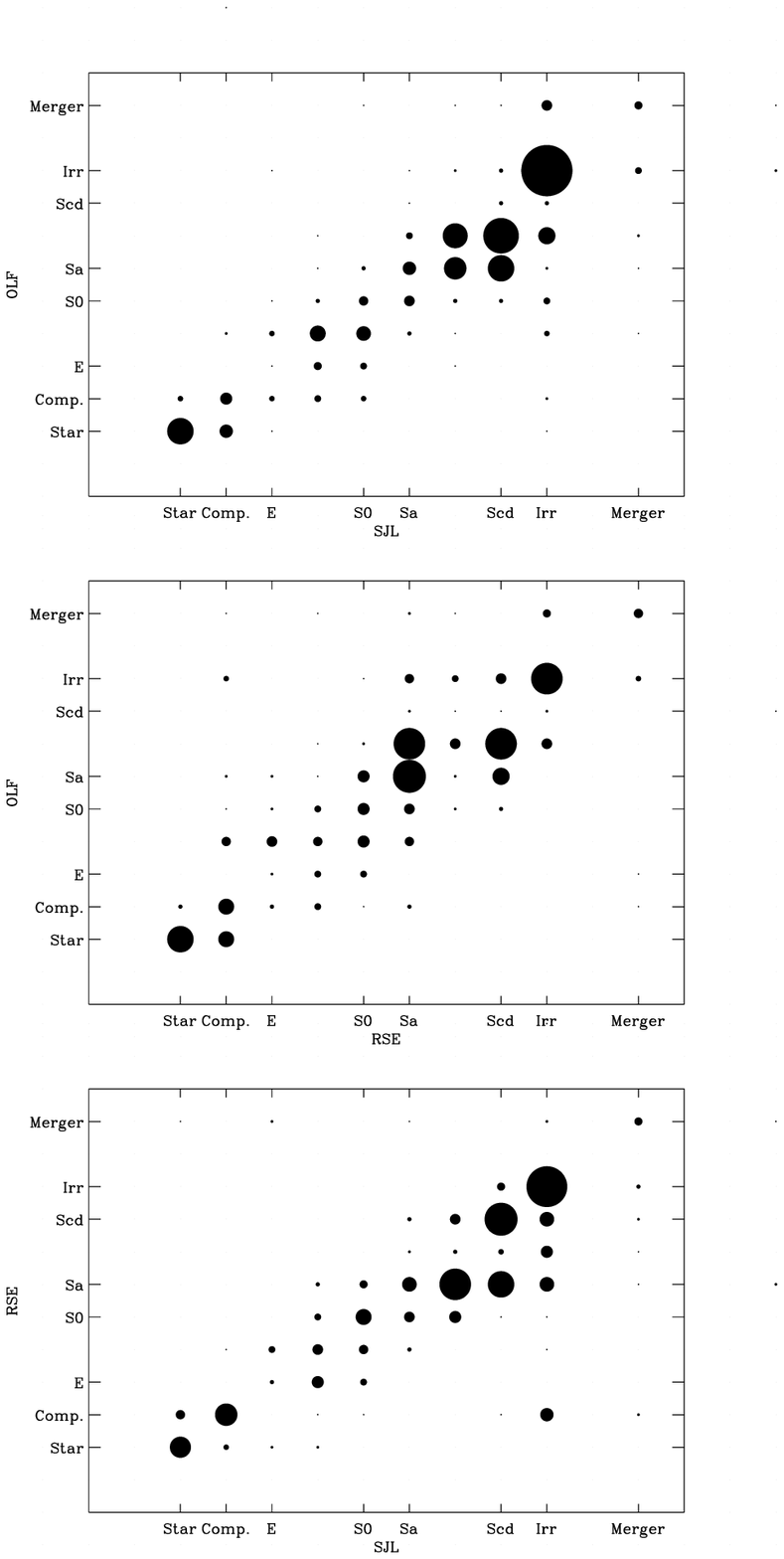}
\psfig{file=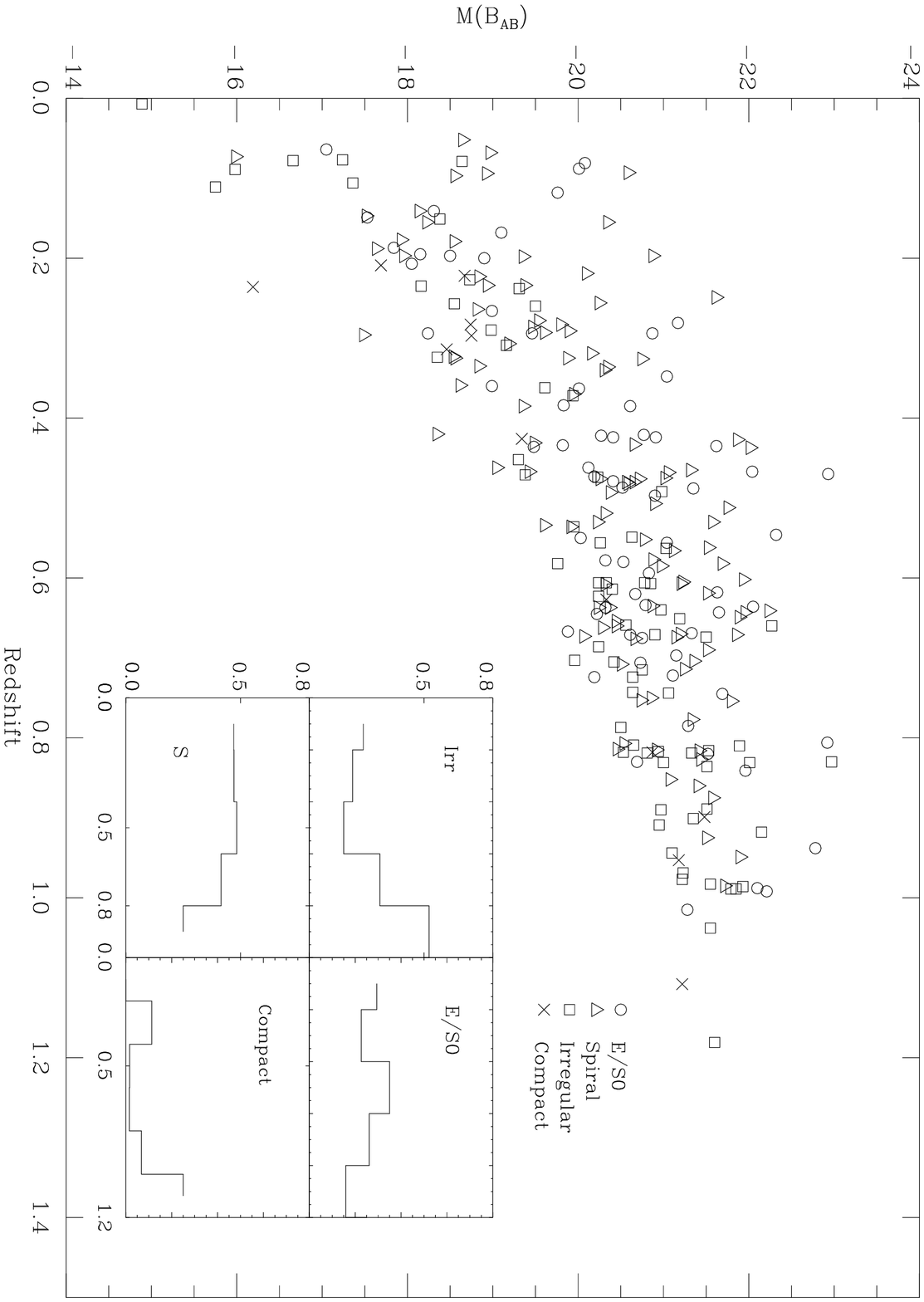,height=0.9\textheight}
\psfig{file=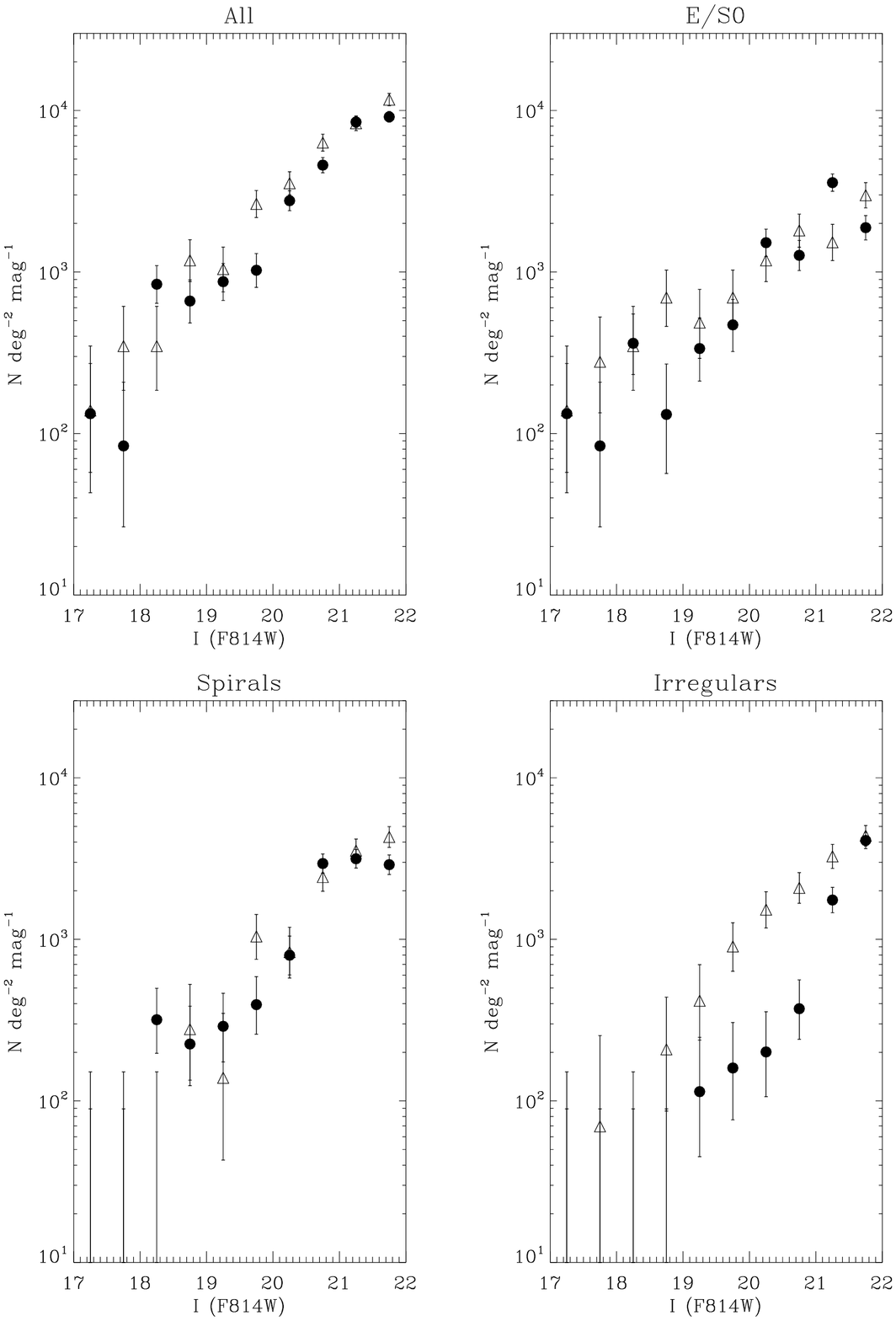,height=0.9\textheight}
\psfig{file=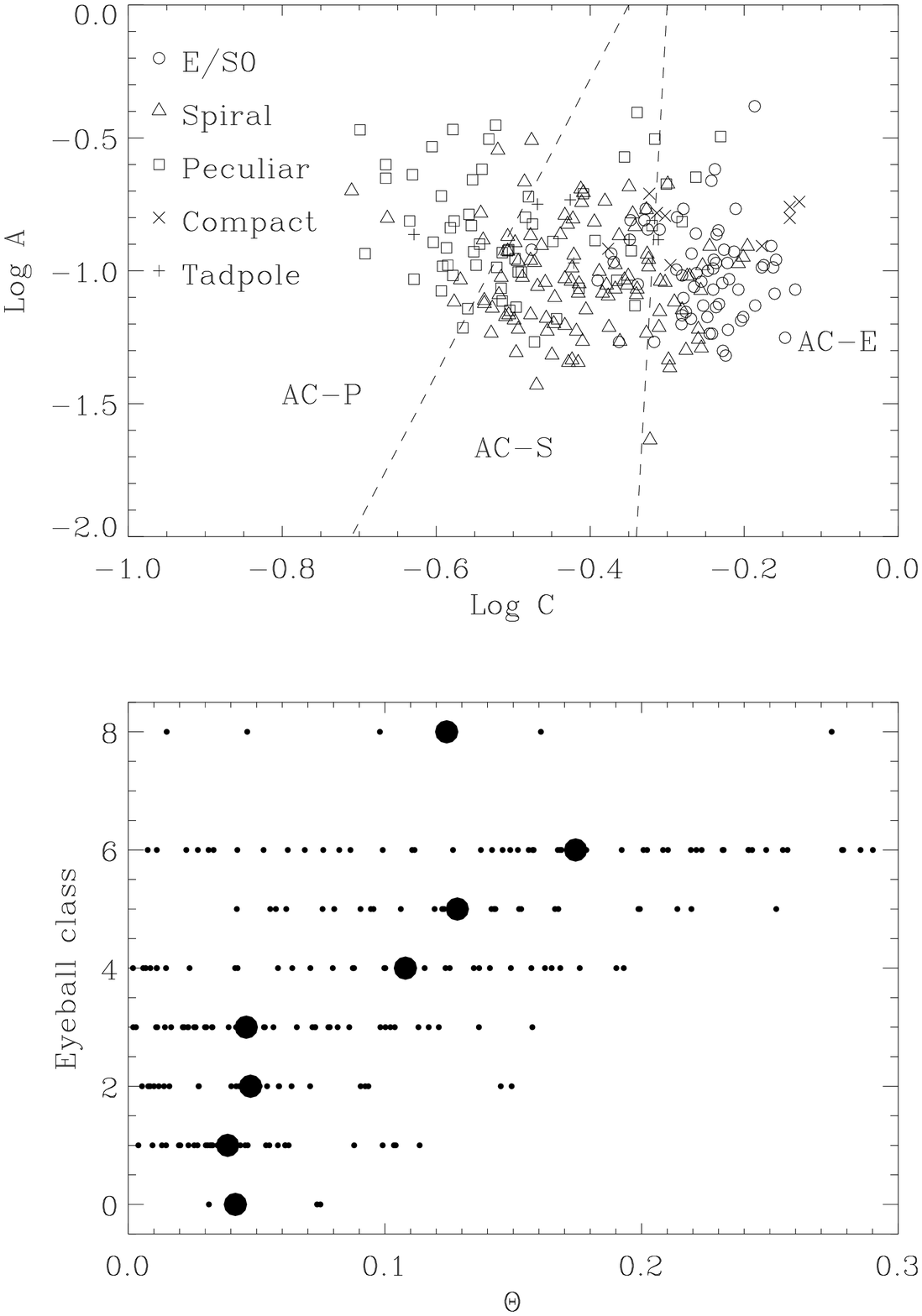,height=0.9\textheight}
\psfig{file=f9.eps,height=0.9\textheight}
\psfig{file=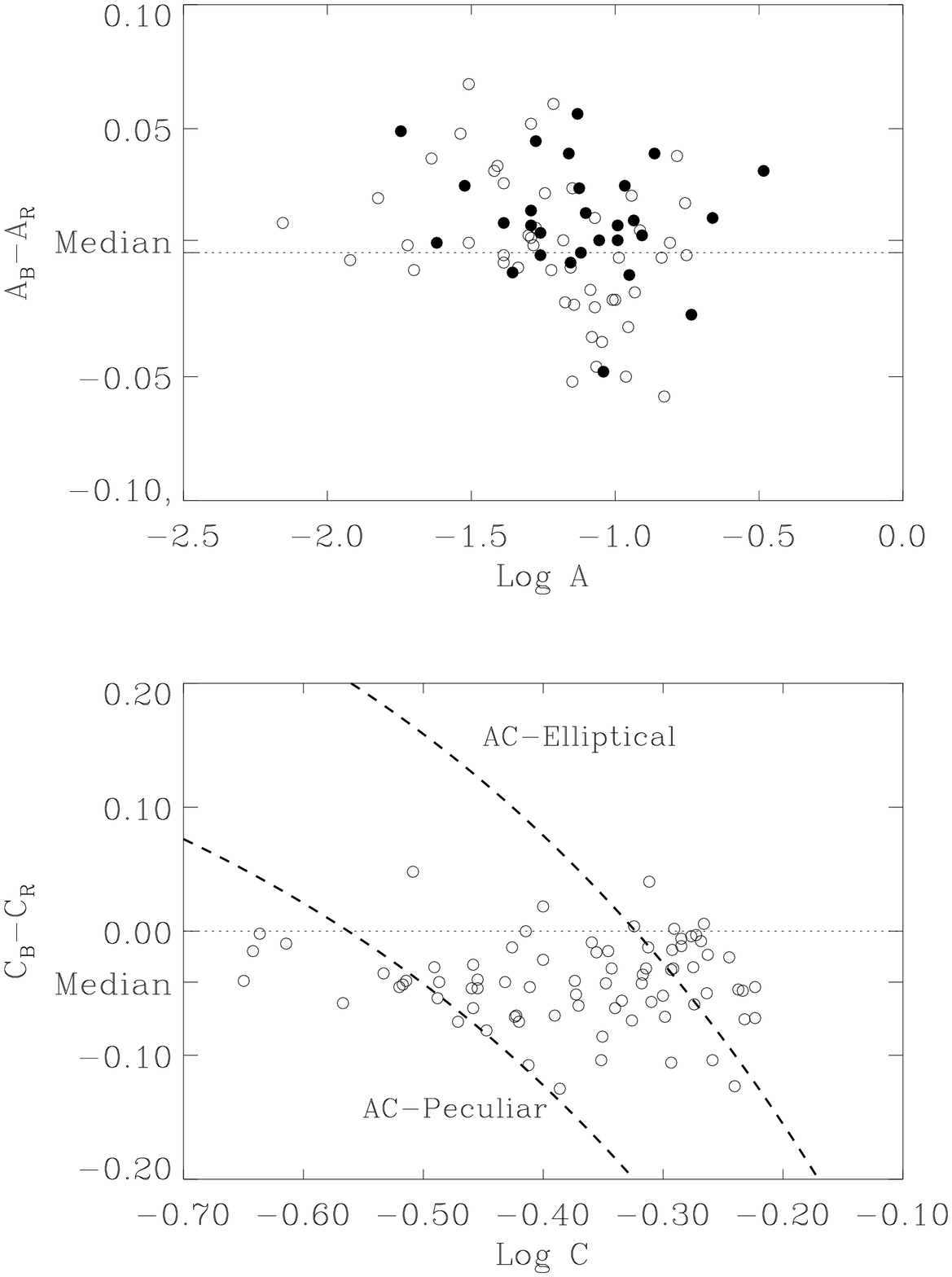,height=0.9\textheight}
\psfig{file=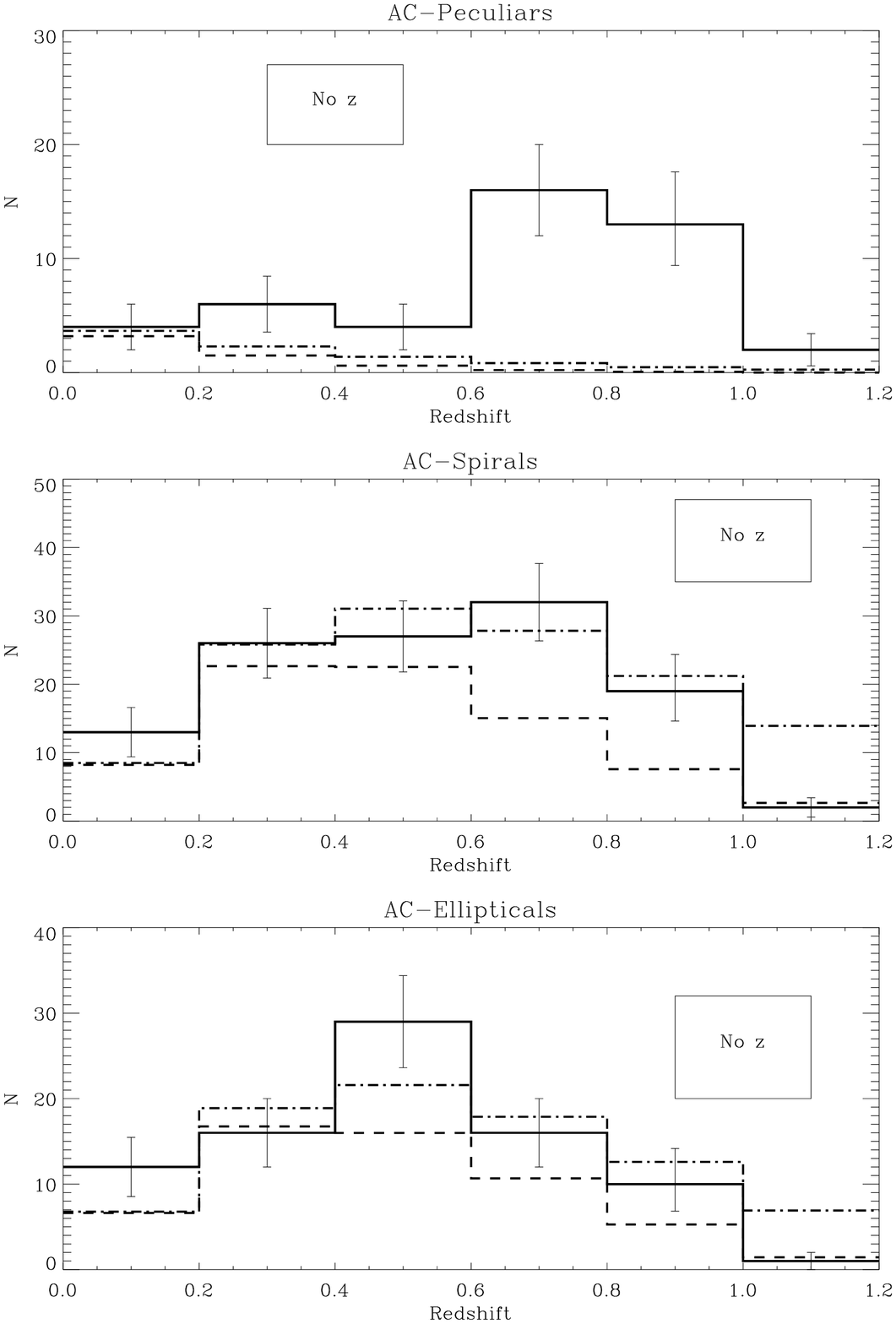,height=0.9\textheight}
\psfig{file=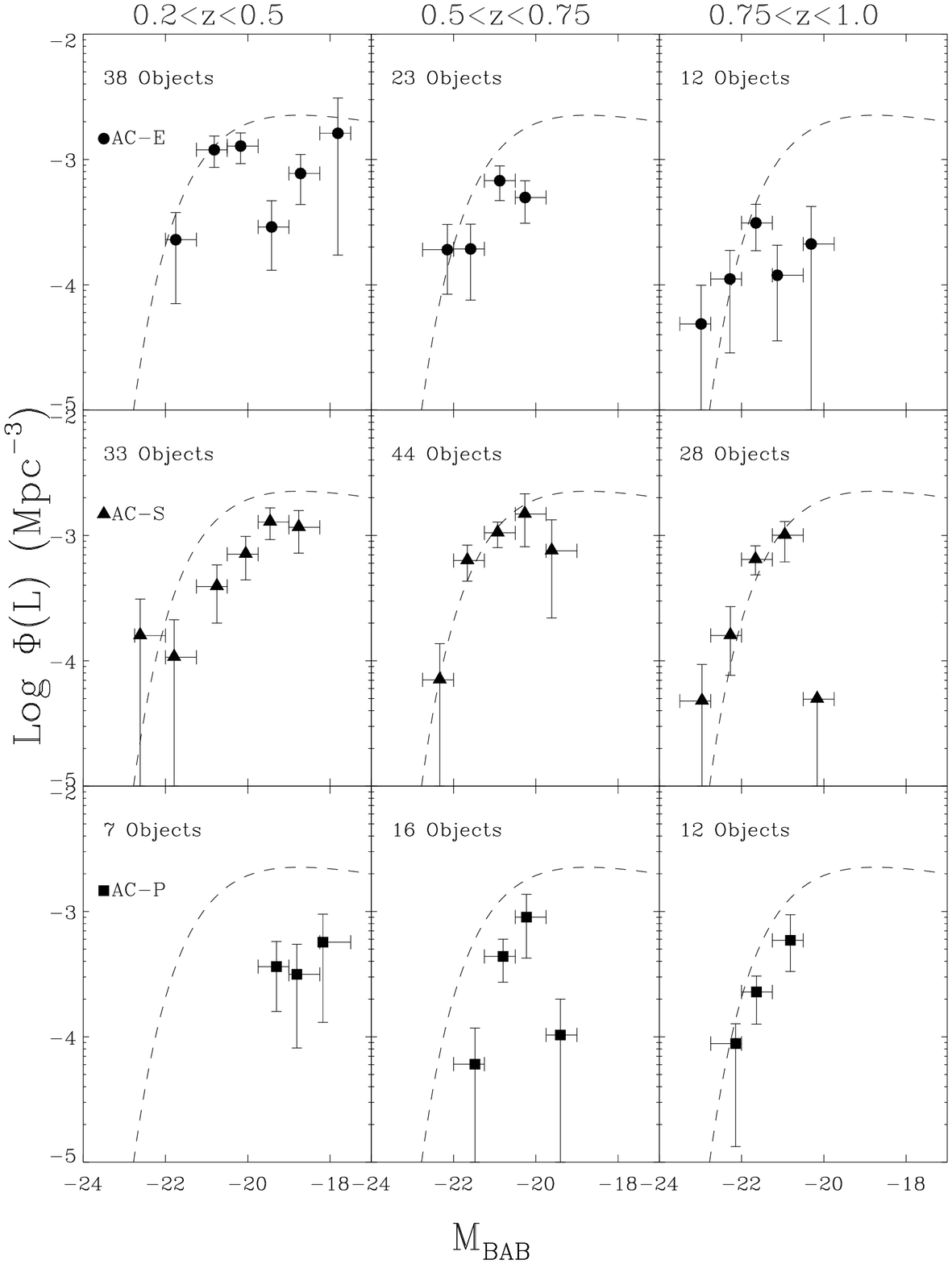,height=0.9\textheight}
\psfig{file=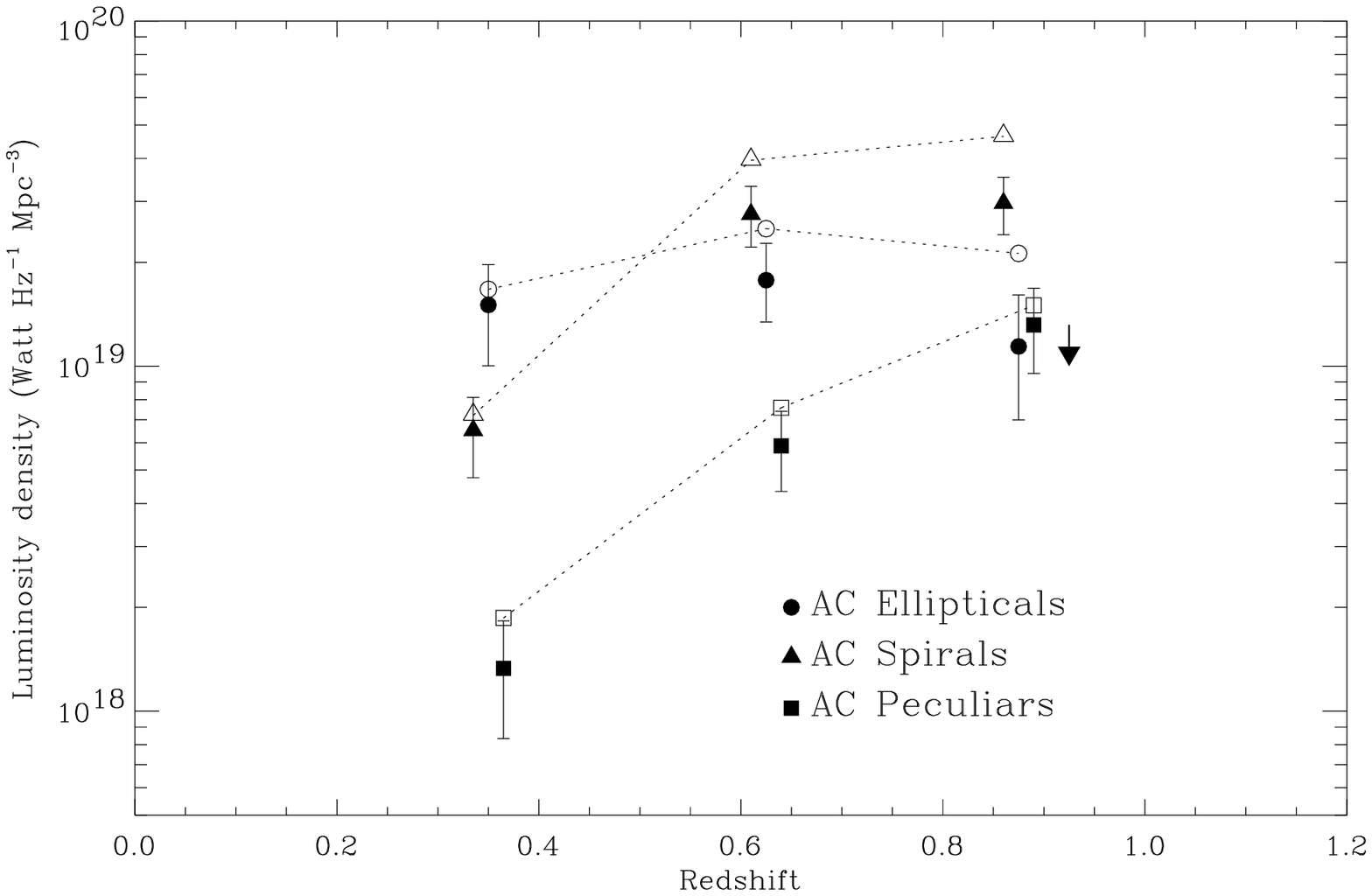,width=0.9\textwidth}
\clearpage
\psfig{file=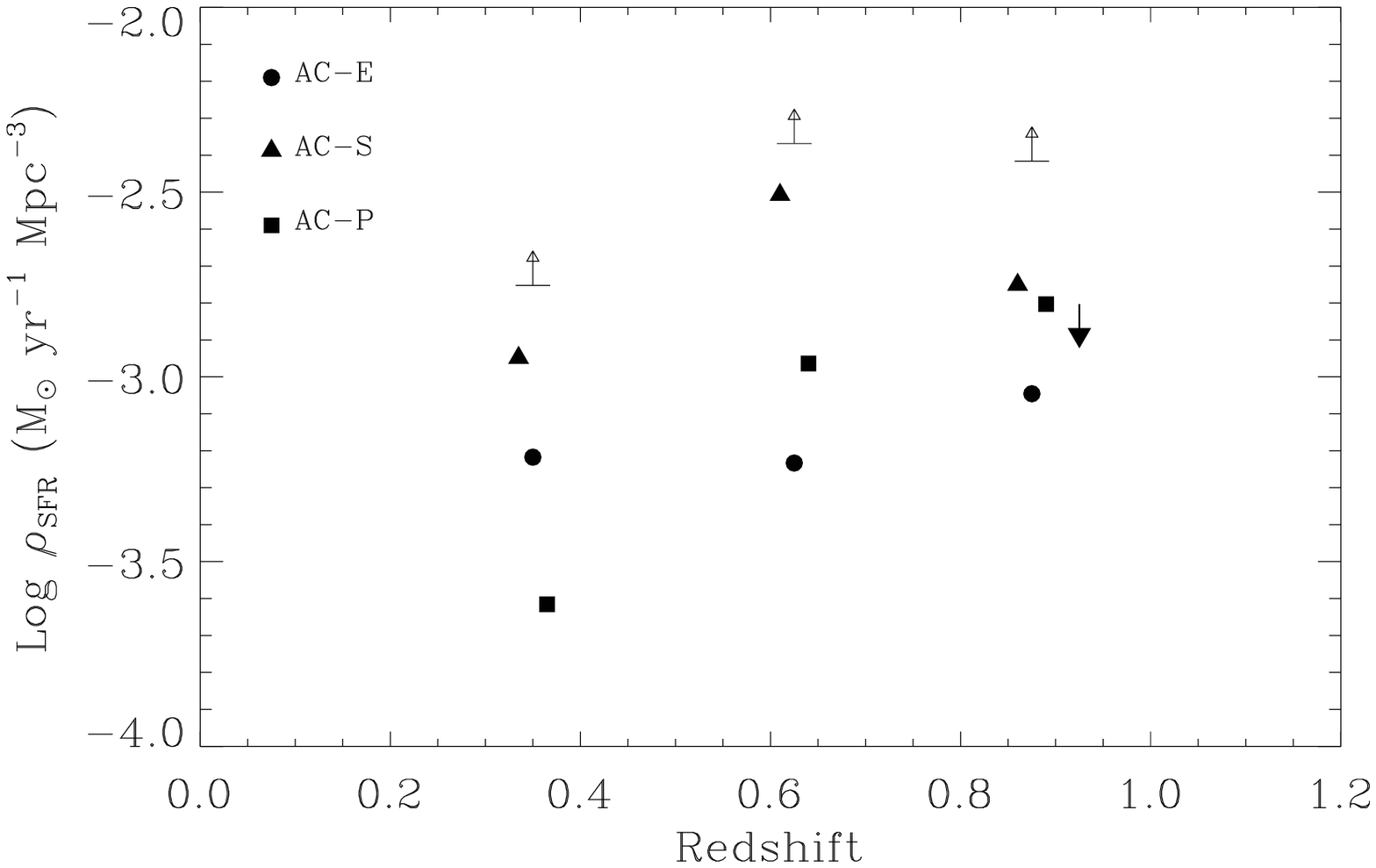,width=0.9\textwidth}

\end{document}